\documentclass[pdflatex,sn-basic,lineno,iicol]{article}
\usepackage[margin=1 in]{geometry}

\usepackage{booktabs}
\usepackage{graphicx}
\usepackage{upgreek}
\usepackage{multicol,multirow}
\usepackage{amsmath,amssymb,amsfonts}
\usepackage{mathrsfs}
\usepackage{amsthm}
\usepackage[figuresright]{rotating}
\usepackage{appendix}
\usepackage[authoryear]{natbib}
\usepackage{ifpdf}
\usepackage[T1]{fontenc}
\usepackage{newtxtext}
\usepackage{newtxmath}
\usepackage{textcomp}
\usepackage{xcolor}
\usepackage{natbib}
\begin{document}

\title{Cycle-to-cycle variations in cross-flow turbine performance and flow fields}

\author{Abigale Snortland\textsuperscript{*1}, Isabel Scherl\textsuperscript{1}, Brian Polagye\textsuperscript{1}, Owen Williams\textsuperscript{2}}

\maketitle
\begin{centering}
\textsuperscript{*}Corresponding author email: abigales@uw.edu

Contributing authors email: ischerl@uw.edu, bpolagye@uw.edu, ojhw@uw.edu

\textsuperscript{*1}University of Washington, Mechanical Engineering, 3900 E Stevens Way NE, Seattle, WA 98195, USA

\textsuperscript{2}University of Washington, Aeronautics and Astronautics, 3940 Benton Lane NE, Seattle, WA 98195, USA

\end{centering}

\begin{abstract}
Cross-flow turbine performance and flow fields exhibit cycle-to-cycle variations, though this is often implicitly neglected through time- and phase-averaging. This variability could potentially arise from a variety of mechanisms -- inflow fluctuations, the stochastic nature of dynamic stall, and cycle-to-cycle hysteresis -- each of which have different implications for our understanding of cross-flow turbine dynamics. In this work, the extent and sources of cycle-to-cycle variability for both the flow fields and performance are explored experimentally under two, contrasting operational conditions. Flow fields, obtained through two-dimensional planar particle image velocimetry inside the turbine swept area, are examined in concert with simultaneously measured performance. Correlations between flow-field and performance variability are established by an unsupervised hierarchical flow-field clustering pipeline. This features a principal component analysis (PCA) pre-processor that allows for clustering based on all the dynamics present in the high-dimensional flow-field data in an interpretable, low-dimensional subspace that is weighted by contribution to overall velocity variance. We find that the flow-field clusters and their associated performance are correlated primarily with inflow fluctuations, despite relatively low turbulence intensity, that drive variations in the timing of the dynamic stall process. Further, we find no evidence of substantial cycle-to-cycle hysteresis. Cycle-to-cycle performance variability occurs earlier in the cycle than flow-field variability, indicating the limits of co-temporal correlation between performance and flow fields, but clustering reveals persistent ties between performance and flow-field variability during the upstream portion of the turbine rotation. The approach employed here provides a more comprehensive picture of cross-flow turbine flow fields and performance than aggregate, statistical representations.
\end{abstract}

\maketitle

\section{Introduction\label{sec:intro}}

Cross-flow turbines are able to harness the kinetic energy in wind, tidal currents and rivers. Relative to axial-flow turbines, cross-flow turbines, referred to a ``vertical-axis'' turbines in the wind sector, operate at lower rotation rates, are insensitive to inflow direction, and may be able to achieve higher power output per unit area within an array \citep{powerDensity}.  However, because cross-flow turbines rotate in a direction perpendicular to the inflow, the blades encounter a continually fluctuating angle of attack and relative inflow velocity that leads to the unsteady, non-linear phenomenon of dynamic stall \citep{Me,Mukul,Mukul2,sebstalldilema,BIANCHINI2016329,Simao,BUCHNER,Dunne}. 
Similar to stall on a static foil, flow separation from the blade is eventually accompanied by a significant loss of lift and increased drag. For cross-flow turbines, dynamic stall severity depends on the ratio of the blade tangential velocity to the inflow velocity -- the dimensionless ``tip-speed ratio''. While cross-flow turbine hydrodynamics are functions of both the blade azimuthal position and the tip-speed ratio, cycle-to-cycle variability in performance and near-rotor flow fields is also observed. This is potentially caused by inflow fluctuations, hysteresis from previous cycles, and the sensitive and stochastic nature of dynamic stall itself. For example, any perturbations in the inflow velocity not only change the kinetic energy available in the flow but also change the instantaneous angle of attack and relative velocity encountered by the blade. This may, in turn, appreciably affect the timing and severity of dynamic stall. Similarly, \cite{Choudhry} hypothesize that dynamic stall is influenced by the state of the boundary layer,  
which suggests that hysteresis from previous cycles, such as the extent of separated flow remaining on the blade at the beginning of the next cycle, may affect future dynamic stall. Additionally, the specific mechanisms underlying cycle-to-cycle variability could depend on the inflow conditions (e.g., Reynolds number), turbine geometry and kinematics.

Cycle-to-cycle variability is commonly neglected in the cross-flow turbine literature. RANS simulations, while often employed to study these flows, are inherently unable to accurately model cycle-to-cycle variations, and the computational expense of LES and DNS precludes their wider use to characterize this variation over a large numbers of cycles \citep{Kuppers}. For experiments, these variations are implicitly neglected when data are time- or phase-averaged. However, phase-averaging can remove information that would otherwise assist in interpreting power production, vortex shedding, and stall, as well as distorting the timing and character of the dynamic stall process by smearing out non-linear phenomena and post-stall events  \citep{Lennie,Riches,Harms,Ramasamy}. As such, more sophisticated techniques that preserve cycle-to-cycle variability could provide additional insight into the flow-field physics. Some cross-flow turbine works do acknowledge the variability that is present in performance and/or flow-field measurements, but simply treat it as experimental uncertainty \citep{Re,MillerSolid,RossScaling,BachantRef,sebtimescales}.

Our objective is to quantify the extent of cycle-to-cycle variation in cross-flow turbine performance and flow fields, its sources, and the relationships between performance and flow-field variability.
To do so, we employ an unsupervised clustering pipeline, comprised of well-developed, data-driven methods, that identifies physically meaningful flow-field clusters with differing dynamics relevant to the dynamic stall process. Clustering was chosen since it can provide a basis for conditionally-averaging experimental data using all the dynamics present, rather than resorting to hand-engineered metrics (e.g., flow-field data based on phase-specific vortex separation from the foil surface). We are also able to correlate these flow-field clusters with performance and investigate the different sources of variability. 
 
While cycle-to-cycle variability has not been explicitly studied for cross-flow turbines, variability in dynamic stall is the subject of several recent works \citep{Harms,TSANG2008129,Lennie,Ramasamy,Ramasamy2021,Kuppers}. 
By using pressure tap measurements on a flapping foil, \cite{Harms} investigated variability in dynamic stall and concluded that the phase-average and a measure of the spread (e.g., standard deviation) were descriptive of the general dynamics for cases where bivariate pressure distributions did not exhibit bimodal behavior. They hypothesized that the cases exhibiting bimodal behavior were more sensitive to inflow conditions and boundary layer unsteadiness. \cite{TSANG2008129} employed a wavelet analysis on lift and drag data from a flapping foil, and showed that non-stationarity between cycles increased with stall severity. They hypothesized that this was the result of non-linear interactions between the fluid force and the pitching motion. \cite{Lennie} used a Covolutional Neural Network on distributed pressure tap measurements to estimate cycle-specific vortex convection speeds, and hierarchical clustering on the coefficient of lift time-series to highlight underlying structure in the cycle-to-cycle variability. They found that vortex convection speed varied cycle-to-cycle, most prominently post-stall, and stall onset differed between the clusters. Because of this, phase-averaging inadequately represented portions of the data set. \cite{Kuppers} showed that using hierarchical clustering on the coefficient of lift time series on a flapping foil produced physically meaningful clusters. They found that one cluster exhibited a higher secondary lift peak than the other even though no clear bimodal behavior existed in the bivariate distributions. \cite{Ramasamy2021} also employed clustering of distributed pressure measurements to investigate cycle-to-cycle variation for a pitching foil. They focused on cases with bimodal behavior in bivariate distributions and found that the clusters diverged in the post-stall region. They also showed that cluster-conditional averages deviated substantially from phase-averages, and that the clusters differed in shedding timing of the dynamic stall vortex (inferred from the pressure data), lift production, and flow recovery. They concluded that the clusters were associated with physical processes.
These studies highlight the presence and complexity of cycle-to-cycle variability in systems that exhibit dynamic stall, and also demonstrate that phase-averaging can generate misleading representations. However, because none of these studies involve flow-field or inflow velocity measurements, they could not quantify the sources of the observed variability, characterize the flow-field variability beyond inference from force/pressure data, or directly correlate force and flow-field variability. 

An analysis that considers both forces and flow fields is necessary to understand the extent and sources of the variability present and the sensitivity of the dynamics. However, unlike forces and pressures, flow-field data is high-dimensional in space and time, which poses apparent limitations on the use of clustering. Specifically, a ``curse of dimensionality'' arises for flow-field data because the algorithmic distances, such as the Euclidian distances, between individual flow-field snapshots begin to converge as the number of data points increases. In other words, from an algorithmic viewpoint, the flow-field snapshots begin to look equally similar and dissimilar from each other even if they differ physically. This makes distinct clusters increasingly difficult to define \citep{beyer}. As such, clustering of high-dimensional data requires adaptations for dimensionality reduction. Principle component analysis, PCA, is a well-known technique useful for decoupling the dynamics of complex, high-dimensional data sets, for feature selection, and for dimensionality reduction \citep{PODbasics,databook,RPCA,Kriegel,Thrun,PODandlift,jetpod}. Additionally, several groups have demonstrated the use of PCA as a means of preserving dynamics otherwise smeared out by phase-averaging of unsteady, vortex-dominated flows \citep{Riches,MullenersRecurrence,Ramasamy}. Clustering in a PCA subspace is particularly useful because the PCA basis optimally maximizes variance in the data \citep{Ramasamy2021}. Because of this, PCA and clustering are commonly used in conjunction \citep{Desoete,Ding,MRIcluster,cloud,bai2017}. Additionally, clusters in a PCA subspace are a useful basis for producing probabilistic reduced-order models and for cluster-based feedback control \citep{KaiserNoack2014,nair2019}. 

We explore the topic of cycle-to-cycle variability in cross-flow turbines using near-blade flow fields and performance (i.e., power output) for two distinct operating conditions. Through this, we characterize the flow-field variability and its connection with turbine performance using unsupervised, hierarchical clustering with a PCA pre-processor on the flow fields. The paper is laid out as follows. Section~\ref{sec:methods} provides a theoretical background for flow-field interactions with the moving rotor, then lays out the methodology for turbine performance and flow-field measurement, flow-field clustering, and correlations between flow-field clusters and performance. Section~\ref{results} quantifies the extent of cycle-to-cycle performance and flow-field variability, then explores how near-blade hydrodynamics, inflow velocity, and hysteresis from previous cycles contribute to the observed variability.

\section{Methods}
\label{sec:methods}
We begin with a discussion of the theoretical blade-level hydrodynamics in Section \ref{theory}, then discuss the experimental acquisition of the simultaneous performance and PIV measurements in Sections \ref{flumedescript}-\ref{piv methods}, and conclude with a detailed description of the PCA analysis and flow-field clustering pipeline in Section \ref{clustering}. A contextual discussion of experimental uncertainty in flow fields and turbine performance is provided in Appendix~\ref{sec:Uncertainty}. The systematic uncertainty (accuracy of the central moment) indicated by this formal analysis exceeds the experimentally-observed variation. However, the aim of this work is to investigate the cycle-to-cycle variability around the central moment that would normally be treated as random uncertainty.

\subsection{Theoretical Blade-level Hydrodynamics and Performance} \label{theory}

\begin{figure}[h!]
\centering
    \includegraphics[width=0.5\linewidth]{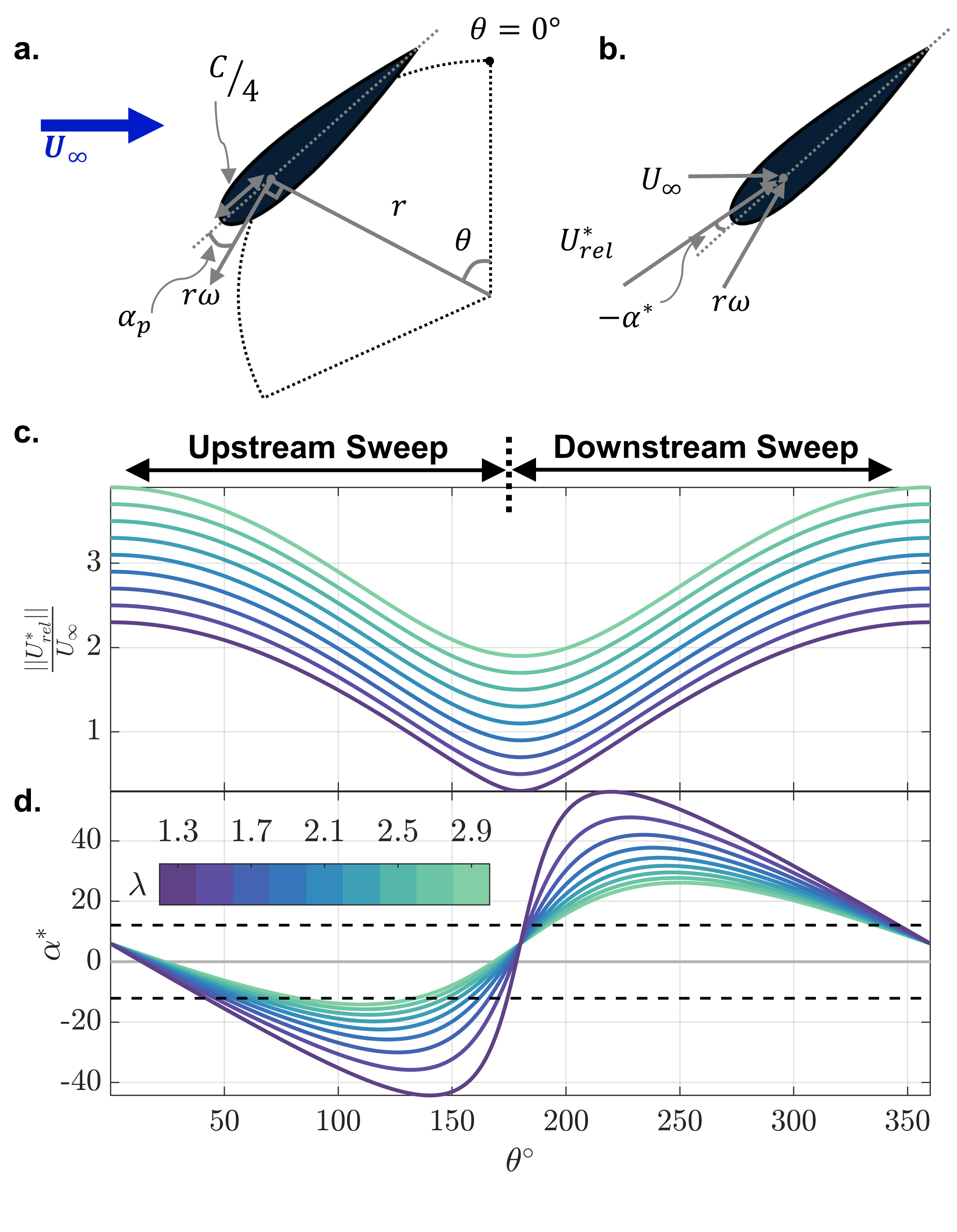}
    \caption{(a,b) Blade geometric definitions, (c) normalized nominal relative velocity trajectories, and (d) nominal angle of attack for the two tip-speed ratios. The tangential velocity is defined tangent to the circular blade path. A positive pre-set pitch angle is depicted in (a), and (b) shows the angle of attack definition. The dashed lines denote the static stall angle in (d) for a foil in rectilinear flow at a similar Reynolds number to the experimental conditions for this study described in Section \ref{turbine methods} ($Re_c\:=\:1.5\times10^5$, \cite{staticalpha}). We note that because of the rapidly varying angle of attack and appreciable induction, the comparison between $\alpha^*$ and the steady-state stall angle is qualitative.}
  \label{nominal}
\end{figure}

To contextualize experimental flow fields and performance, it is instructive to consider the kinematics and dynamic stall theory relevant to the hydrodynamics of cross-flow turbines. A schematic of the blade geometric definitions pertinent to the kinematics is given in Figure \ref{nominal}a,b. Two key factors that govern the near-blade hydrodynamics are the nominal angle of attack, $\alpha^*$ (affecting lift and drag coefficients), and the nominal incident velocity, $U^*_{rel}$, (affecting the magnitude of the lift and drag forces). In the absence of any induced flow (i.e., proximate changes in streamwise and cross-stream velocities as a consequence of interaction with the turbine), the nominal angle of attack, defined as the angle between the chord line and $U^*_{rel}$ at the quarter chord, $C/4$, is
\begin{equation}
\alpha^*(\lambda,\theta)=-tan^{-1}\bigg[\frac{sin(\theta)}{\lambda+cos(\theta)}\bigg]+\alpha_p.
\end{equation}
Here $\alpha_p$ is the blade preset pitch angle, $\theta$ is the blade azimuthal position ($\theta\:=\:0^\circ$ defined as the blade tangential velocity vector pointing directly upstream), and $\lambda$ is the tip-speed ratio. The tip-speed ratio is a non-dimensional representation of the turbine rotation rate, defined as
\begin{equation}
    \lambda\:=\:\frac{r\omega}{U_{\infty}}
\end{equation}
where $r$ is the turbine radius and $\omega$ is the rotation rate.
The nominal incident velocity (relative velocity to $C/4$) is the vector sum of the tangential velocity, $r\omega$, and the freestream velocity, $U_\infty$, such that its non-dimensional magnitude is  
\begin{equation}
	\frac{\|U^*_{rel} (\lambda,\theta)\|}{U_\infty} =\sqrt{\lambda^2+2\lambda cos(\theta)+1}.
\end{equation}
Azimuthal variations in $||U^*_{rel}||$ and $\alpha^*$ over one turbine rotation are shown in Figure \ref{nominal}c,d. Here $\| \|$ denotes the magnitude. For negative $\alpha^*$, on the upstream sweep, the lift vector points inward to the center of rotation and, therefore, the suction side of the blade is the inner surface. Conversely, for positive $\alpha^*$ on the downstream sweep, the suction side of the blade is the outer surface. We refer to these as ``nominal'' quantities because the formulation does not account for induction, the alteration of the inflow by interaction with the turbine rotor. The effect of induction is appreciable but unknown.  

Because $\alpha^*$ and $U^*_{rel}$ depend on $\lambda$, the phase, duration, and severity of dynamic stall are influenced by this parameter. A decrease in $\lambda$ reduces $||U^*_{rel}||$ and increases the range of $\alpha^*$ encountered during a cycle, which corresponds to earlier vortex shedding, increased stall severity, and delayed flow recovery.  
In severe, or ``deep'' dynamic stall cases (lower $\lambda$, larger $\alpha$ ranges), the near-blade flow field is characterized by the formation and roll-up of an energetic dynamic stall vortex that is on the order of the blade chord. This vortex grows to maturity and sheds before the maximum angle of attack is reached. After shedding, the blade experiences a sharp drop-off in lift and an increase in drag. In contrast, any vortex growth in ``light'' dynamic stall cases (higher $\lambda$, smaller $\alpha$ ranges) is prematurely terminated and shedding is induced by downward flow entrainment as $\alpha$ begins to decrease \citep{Mulleners,McCroskey}. Changes to the foil geometry (such as thickness) and the turbine control strategy can change the topology of dynamic stall experienced.

Provided turbine geometry and all other relevant non-dimensional parameters (e.g., Reynolds number, Froude number, blockage) are held constant, phase-averaged hydrodynamic power, $P$, and the global velocity fields, $\vec{\boldsymbol{V}}$, are functions of $U_\infty$, $\lambda$, and $\theta$ \citep{Re,RossScaling}. Within a single turbine cycle, $n$, hydrodynamic power is non-dimensionalized as the coefficient of performance $\eta(\lambda,\theta,n)\:=\:\frac{P(\theta)}{\rho U^3_\infty Lr}$ where $\rho$ is fluid density and $L$ is the blade span. The coefficient of performance is often presented as $C_P$, but $\eta$ is used here for notational simplicity and does not imply a ``water-to-wire'' efficiency. The upstream sweep ($\theta\:=\:0^\circ\:-\:180^\circ$) is commonly referred to as the ``power stroke'' of the turbine as it produces most of the hydrodynamic power, while the downstream sweep ($\theta\:=\:180^\circ\:-\:360^\circ$) is characterized by parasitic drag, post- and secondary-stall events, and boundary layer reattachment. The deviation between nominal and true values for $\alpha$ and $U_{rel}$ (Figure~\ref{nominal}) is most pronounced on the downstream sweep because of momentum extraction by the upstream sweep.

\subsection{Experimental Facility}
\label{flumedescript}
\begin{figure*}[h!]
    \centering
    \includegraphics[width=1\linewidth]{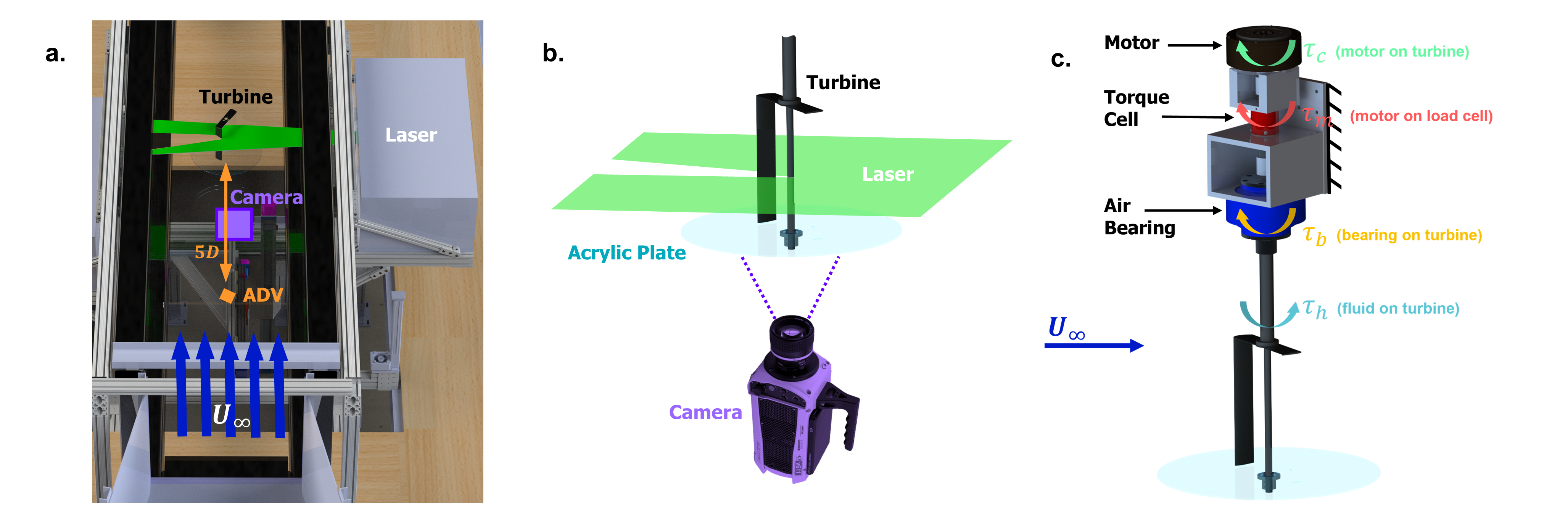}
    \caption{Annotated PIV experimental setup in the flume (a,b) and turbine free body diagram (c). The turbine is one-bladed but utilizes a top strut designed to support two blades.}
    \label{flume}
\end{figure*}

Experiments were performed in the Alice C. Tyler flume at the University of Washington, a rendering of which is shown in Figure \ref{flume}a. The data presented in this paper utilized a mean dynamic water depth, $h$, of 0.5 m, resulting in a channel cross-sectional area $A_C$ of 0.375 m\textsuperscript{2} (0.75 m width). The water temperature was maintained at $36.3\pm0.2$ \textsuperscript{$\circ$}C, giving a $\rho$ of 993.5 kg/m\textsuperscript{3}, and a kinematic viscosity, $\nu$, of $7.1\times10^{-7}$ m\textsuperscript{2}/s. An acoustic Doppler velocimeter (Nortek Vectrino) operating at a 100 Hz sampling rate and positioned approximately 5 diameters upstream of the turbine rotor, measured an average $U_{\infty}$ of 0.7 m/s with a turbulence intensity, $TI$, of 1.8-2.1\% . These conditions corresponded to a Froude number, $Fr\:=\:\frac{U_\infty}{\sqrt{gh}}$, of 0.32 where the gravitational constant $g$ is 9.81 m/s\textsuperscript{2}. 

\subsection{Cross-flow Turbine}
\label{turbine methods}

\subsubsection{Experimental Setup and Measurements}

These experiments utilized a one-bladed (NACA 0018 foil) turbine. The turbine has a radius of 8.6 cm, blade span of 23.4 cm, a blade chord length, $C$, of 4.06 cm and a $6^\circ$ preset pitch. The  support structure comprised of a NACA 0008 foil strut at the top of the blade span and a large acrylic plate (40 cm diameter) at the bottom. The plate facilitates PIV imaging of the mid-span from the cameras positioned below the flume (Figure \ref{flume}a,c). Because turbine torque is measured in line with the drive train, a one-bladed turbine is necessary to directly tie performance variations to the near-blade flow fields (i.e., with a two-bladed turbine, the torque contribution from each blade is ambiguous). The blockage ratio, $\beta\:=\:\frac{2Lr}{A_C}$  is 8.9\% and the Reynolds number,  $Re_C\:=\: \frac{U_{\infty}C}{\nu}$ is $3.95\times10^4$.

As shown in Figure \ref{flume}c, the turbine rotation rate, $\omega$, is regulated by a servomotor (Yaskawa SGMCS-02B servomotor, Yaskawa SGDV-2R1F01A002 drive). In these experiments, we are operating under constant speed control holding $\omega$ constant, which requires the servomotor to apply a variable control torque, $\tau_{c}$, which is measured by a hollow reaction torque cell (Futek FSH02595) rigidly coupled to the motor and flume cross beam. An air bearing (Professional Instruments Company Block-Head 4R low-inertia) absorbs the thrust moment while imparting a minimal bearing torque. 

MATLAB Simulink Desktop Real-Time was used for data collection and turbine control. For each control set point (constant $\omega\:=\:8.8-28$ rad/s in 1.6 rad/s increments, corresponding to $\lambda\:=\:1.1-3.5$ in 0.2 increments), torque cell data were acquired for $> 60$ seconds (sufficient for convergence of the time-averaged statistics) at 1 kHz (upper limit of the acquisition hardware). Blade position was measured by the servomotor encoder with a resolution of $2^{18}$ counts/rotation, also sampled at 1 kHz. The rotation rates were computed by taking the derivative of the measured blade position.

\subsubsection{Blade-level Performance Calculation}
\label{bladeperfcalc}
Hydrodynamic power produced by the turbine is the product of the hydrodynamic torque, $\tau_h$, and $\omega$. At the turbine level, $\tau_h$ is the net hydrodynamic torque produced by the blades, less the parasitic torque from support structure drag. For constant $\omega$ and negligible bearing torque, the torque balance reduces to $\tau_h=\tau_c$.

\begin{figure}[b!]
    \centering
    \includegraphics[width=0.5\linewidth]{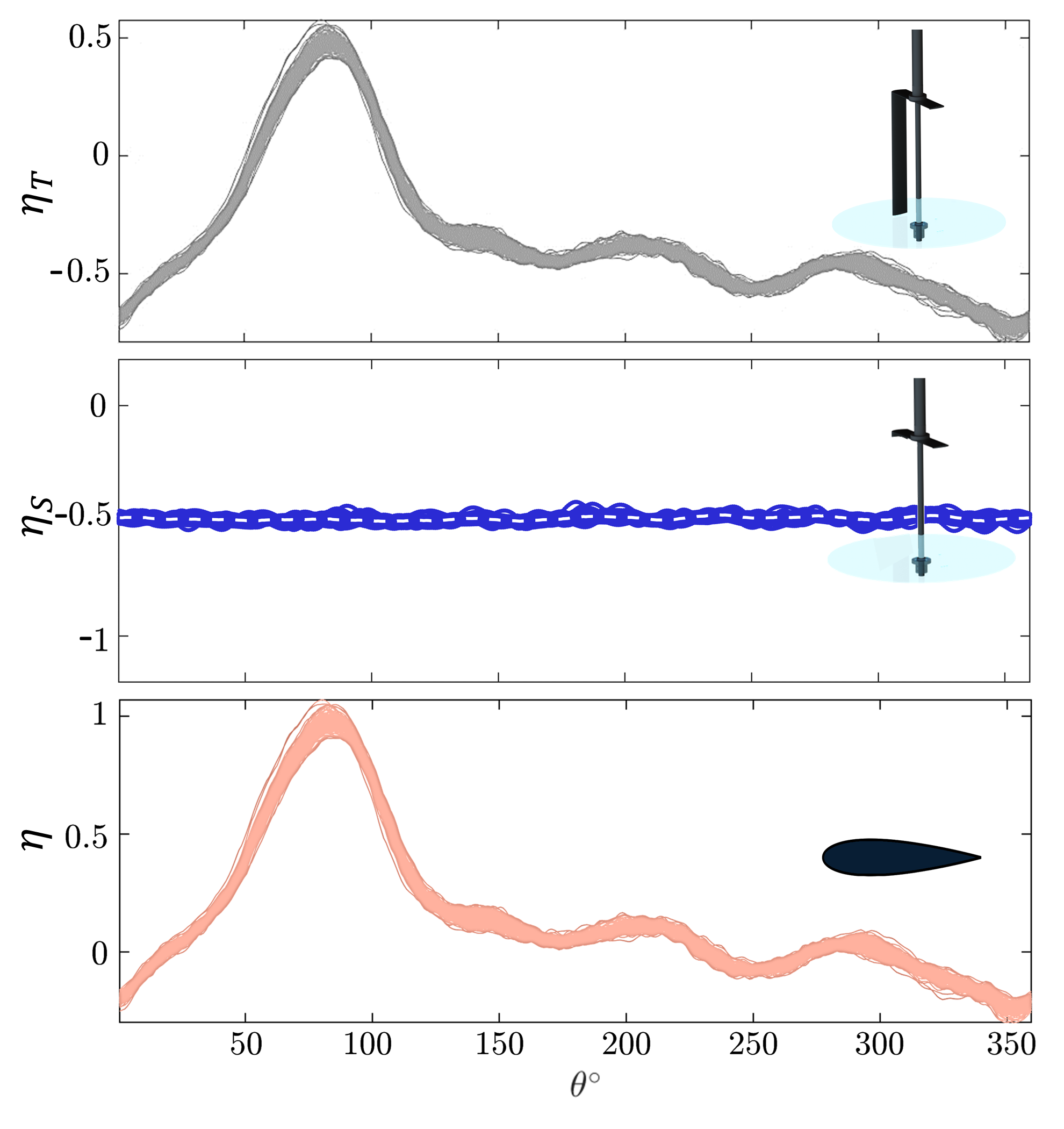}
    \caption{Schematic representations of the calculation of blade level performance, $\eta$, by subtracting support-only ``performance'', $\eta_S$, from the full-turbine performance, $\eta_T$, at the same conditions. $\eta=\eta_T\:- \langle\eta_S\rangle$.}
    \label{bladepower}
\end{figure}

Because the measured torque is dominated by the parasitic torque incurred by large bottom plate, ``full-turbine'' performance is not a meaningful metric. To this end, ``blade-level'' $\eta$ is calculated by subtracting phase-averaged performance for the turbine support structure (no blade), $\langle\eta_S\rangle$, at the same inflow conditions from the full-turbine performance measurements, $\eta_T$ (Figure \ref{bladepower}). Here the $\langle \rangle$ brackets denote the phase-average which is conditional on $\lambda$ and $\theta$, or, in other words, an average for a single operating condition and azimuthal position across multiple cycles. Since the performance data is captured continuously, we utilize a $1^\circ$ $\theta$ bin for phase-averaging. Blade-level $\eta$ is used throughout to describe cycle-to-cycle performance variability. This approach requires that secondary interactions between the blades and support structures are minimal (demonstrated in \cite{stromsupports,hunt}) and that cycle-to-cycle fluctuations are dominated by variation in blade performance and not variation from the support structure. 

All performance measurements were filtered with a low-pass, zero-phase, Butterworth filter to remove high-frequency electromagnetic interference from the servomotor. Both the 75 and 30 Hz cutoff frequencies used for the turbine and support structure performance data, respectively, are $10+$ harmonics faster than the blade passage frequency. Additionally, the Butterworth filter has no effect on the time-averaged performance. Therefore, it unlikely the filter is removing any hydrodynamic power. 

\begin{figure}[t!]
    \centering
    \includegraphics[width=0.5\linewidth]{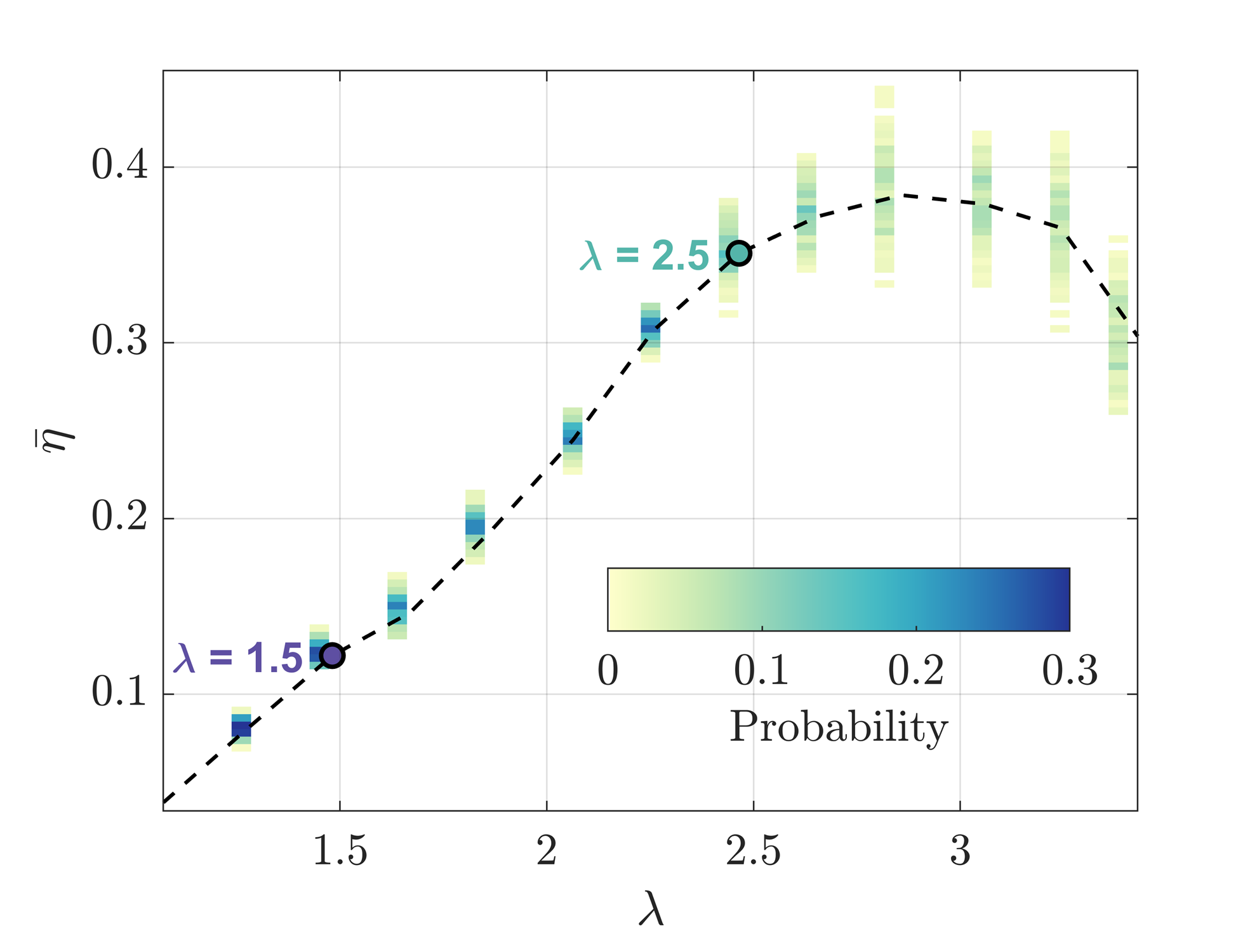}
    \caption{Time-averaged coefficient of performance for the one-bladed turbine investigated. The dashed line represents the time-average at each $\lambda$. Near-optimal (green dot) and sub-optimal (purple dot) cases are highlighted. The histograms (yellow to blue shading) describe the range of time-average performance over individual cycles at each $\lambda$.}
    \label{timeaverage}
\end{figure}

In calculating $\eta$, we note there is some ambiguity in the choice of $U^3_\infty$. For example, one could attempt to use instantaneous velocity measurements, but there is a temporal mismatch between the time the freestream velocity is measured and when it interacts with the rotor. Additionally, we cannot use PIV to estimate inflow since our field of view does not extend far enough upstream to measure the undisturbed inflow. Our notional choices are (1) to use a time-average of all of the cubed freestream velocity measurements acquired at a tip-speed ratio set point, or (2) apply an advection correction, as in \cite{Briancontrol}, to compute the instantaneous freestream velocities and calculate cycle-specific kinetic power. The advection correction is based on a cross-correlation of $U_{\infty}$ with measured power and the application of Taylor's frozen flow hypothesis. Option (2) performs poorly in these experiments, in that it generates greater relative variability in $\eta$ than measured power, possibly because of the single-bladed configuration. Consequently, we utilize option (1), but in Section \ref{freestreamimpact}, we employ a hybrid approach to estimate ``cluster-specific'' kinetic power as an explanatory factor for cycle-to-cycle performance variation. A characteristic time-averaged performance curve, utilizing option (1), is given in Figure \ref{timeaverage}.

\begin{figure}[h!]
\centering
    \includegraphics[width=0.5\linewidth]{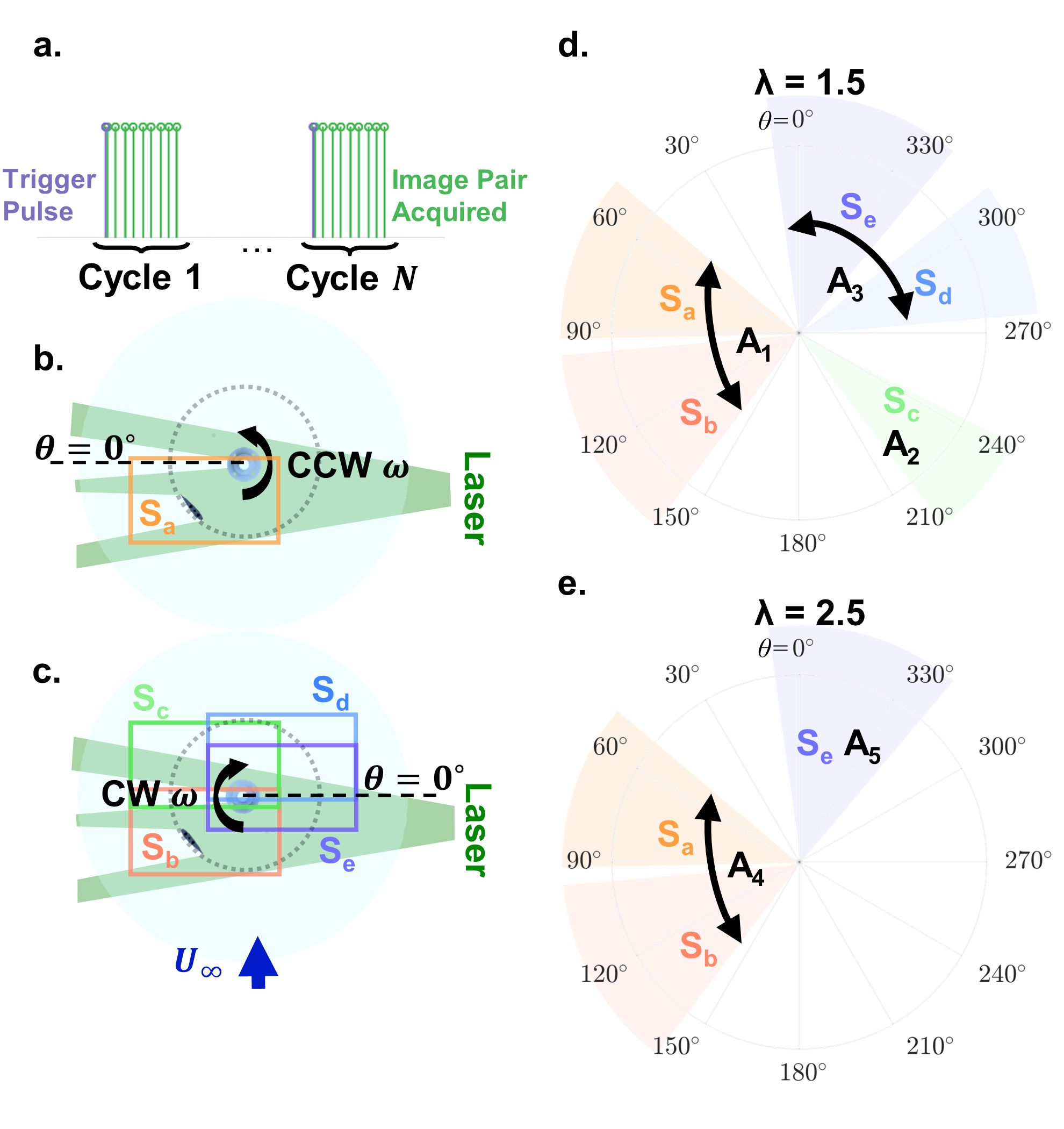}
    \caption{(a) A simple timing diagram for the PIV acquisition. Acquisition begins each cycle by the trigger pulse and then a pre-determined number of phase-locked image pairs are aquired at the prescribed camera capture frequency, For example, in $S_a$, 9 image pairs are taken after the trigger pulse each cycle. Field of view locations captured for (b) CCW rotation and (c) CW rotation. Example blade and support structure shadows are shown. Note that the $\theta = 0^\circ$ reference changes with rotation direction, such that $S_a$ does not overlap with $S_b$. (d,e) PIV segments captured for the different fields of view and visual representation 
 of corresponding $A$ matrix assignments for input into the PCA pre-processor.} 
   \label{segsbreakdown}
\end{figure}

\subsection{Flow Fields}
\label{piv methods}

\begin{table}
\begin{centering}
\caption{PIV parameters for $\lambda\:=\:1.5$ and $\lambda\:=\:2.5$.}
\label{piv exp params}
\begin{tabular}{ l r|c|c }
\toprule
\multicolumn{2}{c}{Parameter} & $\lambda\:=\:1.5$ & $\lambda\:=\:2.5$\\
\midrule
$\omega$ & Hz & 1.9  & 3.2  \\
$\Delta\theta$ $(S_a)$ & & $5.1^\circ$ & $5.1^\circ$ \\
$\Delta\theta$ $(S_b,S_e)$ & & $6^\circ$ & $6^\circ$ \\
$\Delta\theta$ $(S_c,S_d)$ & & $6.6^\circ$ & N/A \\
Camera freq. $(S_a)$ & Hz & 135 & 225 \\
Camera freq. $(S_b,S_e)$ & Hz & 115 & 191.7 \\
Camera freq. $(S_c,S_d)$ & Hz & 105 & N/A  \\
Total image pairs $(S_a,S_b,S_e)$ & & 1251 & 1251 \\
Total image pairs $(S_c,S_d)$ & & 834 & N/A \\
\bottomrule
\label{PIV table}
\end{tabular}
\end{centering}
\end{table}

Two-dimensional, two-component, phase-locked, flow-field measurements were obtained in a streamwise plane at the mid-span of the turbine, simultaneously with turbine performance measurements for two tip-speed ratios: $\lambda\:=\:1.5$ (sub-optimal performance) and $\lambda\:=\:2.5$ (near-optimal performance). While the performance measurements were continuously recorded over the entire turbine rotation, each  PIV acquisition only captured flow-field data during each cycle at discrete, prescribed $\theta$ positions. PIV acquisition was controlled by TSI Insight, and acquisition for each cycle commenced upon receipt of trigger pulses sent at a specified $\theta$ from the Simulink model controlling the turbine. The PIV system returned pulses at each image pair capture instance (Figure \ref{segsbreakdown}a), the timing of which were logged by the Simulink model. Using these timing signals, the PIV snapshots were aligned with performance in post-processing. We note that this does not produce perfect phase-locking, but ``phase jitter'' between image pairs at the same $\theta$ was on the order of $0.009^\circ$, which we deem insignificant for the current analysis.

The general arrangement of the PIV laser and cameras is shown in Figure \ref{flume}a. A dual cavity, Nd:YLF laser (Continuum Terra PIV) capable of a repetition rate of 10 kHz, illuminated the flow with an approximately 2 mm thick light sheet in the field of view, FoV. A high-speed camera  (Vision Research Phantom v641) with 2560 x 1600 resolution and a 105 mm lens at f\# 16, and a calibration of 11.67 pixels/mm resulted in a FoV of 21.9 x 13.7 cm [$5.4C$ ($1.3D$) x $3.4C$ ($0.8D$)]. With this magnification and the flow seeding (10 $\mu$m hollow-glass beads), the resulting particle images were approximately 2-3 pixels in diameter. To minimize laser reflections at the blade surface, matte black paint was applied. 

As shown in Figure \ref{segsbreakdown}b,c, the combination of the camera lens focal length and streamwise extent of the laser sheet necessitated multiple FoVs to capture the important stages of dynamic stall and flow recovery. Each FoV outline is color-coded to denote the flow segment, $S_k$ (where $k$ denotes the segment letter), used in the clustering analysis detailed in Section \ref{clustering}. A single flow segment is captured per PIV experiment. The PIV parameters specific to each flow segment are listed in Table \ref{PIV table}. Sequences of 9 image pairs for $S_a$, $S_b$ and $S_e$, and 6 image pairs for $S_c$ and $S_d$  were acquired per rotational cycle with prescribed angular displacements, $\Delta\theta$, between frames ranging from $5.1 - 6.6^\circ$, depending on desired phase resolution. The phase resolution is adjusted by setting the camera capture frequency equal to $\frac{\omega}{2\pi \Delta\theta}\times 360$. A total of 139 image pairs ($N\:=\:139$ turbine cycles, limited by camera storage capacity) were acquired at each phase position, $\theta$. FoV positioning relied on a combination of camera and turbine movement. A motorized, three-axis gantry positioned the camera relative to the laser sheet and provided the dominant adjustment for cross-stream FoV positioning, as well as fine adjustments in the streamwise direction. The limited streamwise extent of the laser sheet (Figure \ref{segsbreakdown}b,c) necessitated shifting the turbine by $\approx\frac{1}{2}D$ upstream to illuminate and capture the downstream blade sweep. 
Both the turbine shaft and blade cast shadows in the laser sheet. Therefore, to obtain data adjacent to the suction and/or pressure sides of the foil at all phases of interest, PIV measurements were repeated with the turbine spinning in both clockwise ($S_b\:-\:S_e$) and counter-clockwise ($S_a$) directions as depicted in Figure \ref{segsbreakdown}b,c. By changing the direction of rotation, we exploit the symmetry of the system to minimize the impact of blade shadows on the following analysis. In this way, we were able to image the suction side of the blade throughout the upstream sweep. For the downstream sweep, we choose to capture the side of the blade with the more visually distinct dynamics.

PIV processing was performed in LaVision DaVis (version 10.1.1). Background subtraction using a Butterworth filter on phase-matched images mitigated residual reflections and background illumination variation. The shadowed regions, visible turbine structures, and remaining reflections were manually masked prior to PIV processing. These masks are specific to each phase, as the shadows and location of the turbine structures vary with blade position. These phase-specific masks applied are functions of $S_k$, $\lambda$, and $\theta$, but constant for all $n$. Processing utilized a multi-grid, multi-pass cross-correlation algorithm with adaptive image deformation with a 75\% overlap and 32 x 32 pixel final interrogation window size resulting in a 0.69 mm vector spacing (approximately 60 vectors per blade chord). Spurious vectors (less than 2\%) were removed with a universal outlier median filter utilizing a 9x9 filter region and a threshold value of 1.5 for the $\lambda\:=\:1.5$ cases and 2.5 for the $\lambda\:-\:2.5$ cases. 

\begin{figure*}[h!]
    \centering
    \includegraphics[width=1\linewidth]{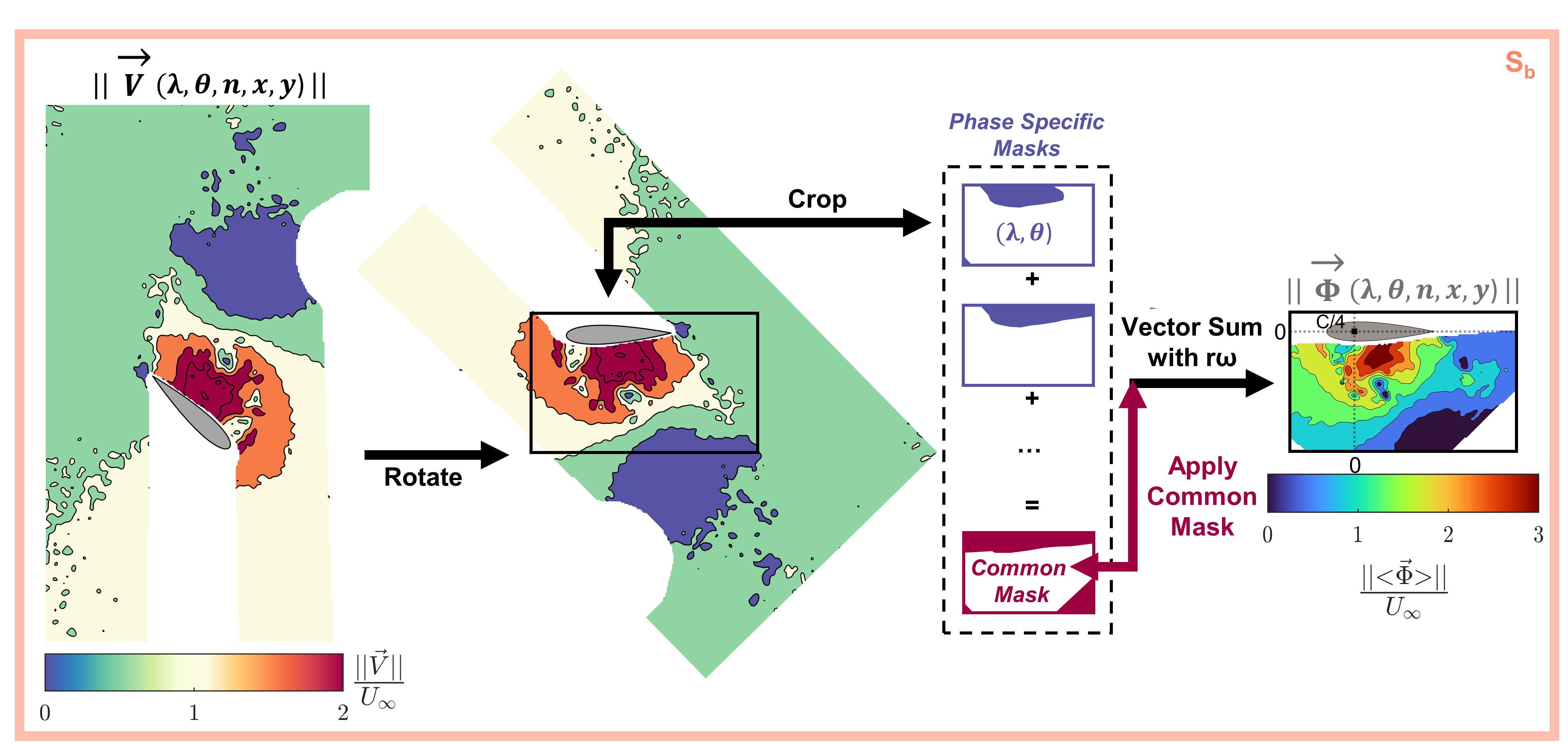}
    \caption{Overview schematic of the flow-field rotation pipeline. The center of rotation between the different fields of view is aligned, then the fields are rotated and cropped into a common blade-centric reference frame. A common mask between all the fields of view in a flow segment is applied and the velocity fields relative to the blade are calculated.}
    \label{croprot}
\end{figure*}

All vector field post-processing was performed in MATLAB (R2020b). To align the different flow segments and limit the contribution of blade rotation in the PCA pre-processing, each FoV was translated from the flume reference frame to a blade reference frame. Figure \ref{croprot} provides an overview of this process. First, we located and aligned the center of rotation between the different FoVs. Tracking a small dot on the end of the shaft and registering all resultant PIV vector fields to a common shaft location corrected phase-dependent shaft procession (slight run-out on the cantilevered turbine shaft). Because of parallax and differences in the index of refraction between air and water, the imaged center of the turbine shaft does not correspond to the center of rotation of the turbine at the imaging plane. Therefore, we determined a best-fit location of the center of rotation in each FoV by manually fitting blades to the masked regions. Second, with the center of rotation located, the flow fields were rotated to the blade-centric coordinate system by locating the turbine blade in each $\vec{\boldsymbol{V}}$ field and rotating the entire field to a common blade position. Third, after rotation, a constant crop boundary and common, segment-specific, mask were applied to each frame, and the relative velocity fields with respect to the blade were computed as the vector sum of $\vec{\boldsymbol{V}}$ with the blade tangential velocity component, $r\omega$. The segment-specific mask was a function of $\lambda$ only and was formed by combining all the phase-specific masks in a specific segment. Finally, the cropped fields were interpolated to a common grid relative to $C/4$. The cropped relative velocity fields, $\vec{\boldsymbol{\Phi}}(\lambda,\theta,n,x,y)\:=\:[u_{rel}(\lambda,\theta,n,x,y)\:,\:v_{rel}(\lambda,\theta,n,x,y)]$ within each $S_k$ are functions of, $\lambda$, $\theta$, and $n$. 

\subsection{Correlating Turbine Performance and Flow-field Variability with Hierarchical Clustering }
\label{clustering}

Cycle-to-cycle variations are present in performance and flow fields, but the correlation between the two is not obvious \textit{a priori}. For example, if flow fields are observed in isolation, the significance of the observed flow structures on turbine performance is unknown. Throughout this work, we use ``correlation'' in a qualitative sense as we do not formally compute correlation coefficients. The flow fields are high-dimensional and a reduced order representation of $\vec{\boldsymbol{\Phi}}$ is needed to compactly describe cycle-to-cycle variation. Clustering with a PCA pre-proccesor allows us to do so while considering all of the flow-field dynamics (and the interplay between them) in an unsupervised manner without relying on hand-engineered metrics. Here we describe the flow-field clustering pipeline used for each flow segment. Using this pipeline, we correlate the variability between the simultaneously captured performance and flow fields. 

\subsubsection{Flow-field PCA Pre-processor}
\label{piv pca}

\begin{figure*}[h!]
    \centering
    \includegraphics[width=1\linewidth]{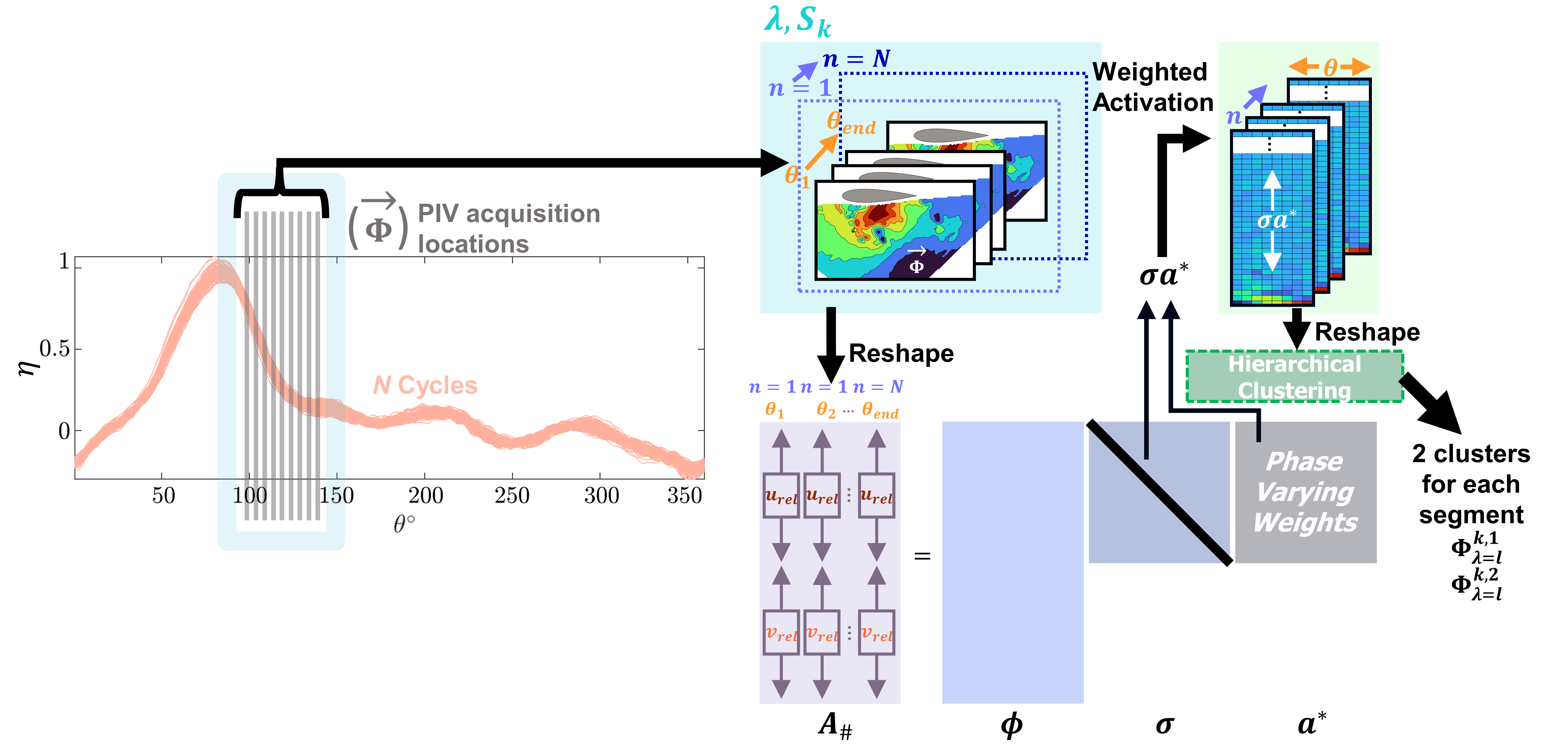}
    \caption{Overview of the flow-field clustering pipeline. Everything upstream of the hierarchical clustering is considered an element of the PCA pre-processor. This involves performing a singular value decomposition on the relative velocity fields in the blade-centric reference frame. The weighted activations (multiplication of the singular values and the phase-varying weights) are then separated by cycle. For each flow segment, the hiearchical clustering algorithm identifies two clusters from this population of cycles. }
    \label{clustering pipeline}
\end{figure*}

Figure \ref{clustering pipeline} describes the flow-field clustering pipeline. Here the PCA pre-processor enables clustering of all the dynamics using a low-dimensional subspace that is interpretable and weighted by contribution to overall velocity variance.
\newline

\textit{\textbf{PCA Setup:}} PCA represents the dynamics of complex data sets as the linear summation of orthogonal modes ranked by the amount variance in the data they describe \citep{PODbasics}. The singular value decomposition (SVD) is used to compute the PCA modes for a data matrix $\boldsymbol{A}_\#$ as
\begin{equation}\label{Eq:SVD}
\boldsymbol{A}_\#=\boldsymbol{\phi}\boldsymbol{\sigma}\boldsymbol{a}^*.
\end{equation}
where the PCA modes are columns of $\boldsymbol{\phi}$, the singular values, which quantify the contribution of each mode, are diagonals of $\boldsymbol{\sigma}$, and the phase evolution of the modes make up the rows of $\boldsymbol{a}^*$. Here $*$ denotes the matrix transpose and $\#$ specifies a specific data matrix. The percentage of the flow-field variance described by a specific mode, $\phi_j$, is quantified using the corresponding singular value $\sigma_j$, as $\Psi_j\:=\:\frac{\sigma_i}{\sum_{j = 1}^{J}{\sigma_j}}$ where $J$ is the total number of modes (smallest dimension of $\boldsymbol{A}_\#$).  

Here, each $\boldsymbol{A}_\#$ is made up of reshaped $\vec{\boldsymbol{\Phi}}$ column vectors, $u_{rel}$ stacked on $v_{rel}$. Any missing values (from the removal of spurious PIV vectors during processing) are linearly interpolated. A common mask is defined for each $\boldsymbol{A_\#}$ as the union of the segment-specific masks. Any data in the common mask region is completely removed ahead of the PCA processing. The presence of the shadowed regions meant a common mask across all $\theta$ (sum of all the segment-specific masks) would occlude the majority of the FoV from analysis, so different segments, $S_k$, at each $\lambda$ were strategically grouped to maximize data yield. This results in five distinct PCA computations of $\boldsymbol{A_\#}$ (Figure \ref{segsbreakdown}). The columns of each $\boldsymbol{A_\#}$ were sorted chronologically [$(n=1,\theta_1),(n=1,\theta_2)...(n=N,\theta_{end})$] with the row-wise mean subtracted prior to SVD decomposition. 
\newline

\textit{\textbf{Computing Weighted Activation Pictures:}} To compactly describe how the dynamics evolve within a rotation, ``weighted activation pictures'' in the PCA subspace are constructed as $\boldsymbol{\sigma}$$\boldsymbol{a}^*$. This multiplication may also be interpreted as the projection of each $\vec{\boldsymbol{\Phi}}$ onto the PCA basis, and encodes both the dominance and phase evolution of all the dynamics (PCA modes) of the system into lower-dimension representations for the hierarchical clustering algorithm. The result of this multiplication is 139 weighted activation pictures (one for each $n$) in each $S_k$. Each ``picture'' is composed of column vectors of the individual weighted activation profiles across all of the computed PCA modes. The individual weighted activation profiles are functions of $\theta$, and describe the dominance of each mode across the range of $\theta$ included in each $S_k$. The weighted activation profiles associated with a mode, $\phi_j$, are denoted as $a_j$.

\subsubsection{Flow-field Clustering}
\label{piv cluster}
After the PCA pre-processor, each $\vec{\boldsymbol{\Phi}}$ field is represented in a low-dimensional form suitable for clustering. While a few of the $\boldsymbol{A_\#}$ contained multiple $S_k$ (i.e., common PCA modes were calculated for segment pairs as shown in Figure \ref{segsbreakdown}), each $S_k$ was clustered separately after PCA pre-processing. This approach was taken because each $S_k$ is an independent data set (e.g., $n=1$ in $S_a$ is unrelated to $n=1$ in $S_b$). The weighted activation pictures in each $S_k$ (each weighted activation picture encoding the flow-field dynamics over 1 cycle), were reshaped into column vectors and sorted into one of two clusters via a hierarchical clustering algorithm utilizing a Ward linkage method (employing a Euclidian distance metric) to minimize variance within a cluster \citep{ward1963}. We utilized the implementation within the MATLAB ``clusterdata'' function. From this, each of the 139 turbine cycles for each flow segment are associated with one of two clusters. 
Two clusters were prescribed because this corresponded to the maximum number of clusters that produced unique cluster-averaged weighted activation profiles (phase average of all profiles within a cluster). In other words, if more than two clusters were prescribed, two or more of the cluster-averaged profiles were almost identical.

\subsubsection{Cluster Assignments}
\label{clusterassign}
After flow-field clustering, we have a set of cluster assignments in each $S_k$ (Figure \ref{clustering pipeline}). We can use these cluster assignments to evaluate ties between the flow fields, $\vec{\boldsymbol{\Phi}}$, and performance, $\eta$, associated with the $\vec{\boldsymbol{\Phi}}$ clusters. This basis also allows us to investigate ties between the performance and flow-field variability and the potential sources of this variability by 
considering both conditional-averages and individual trajectories of different parameters based on flow-field cluster assignment. As shown in Figure \ref{clustering pipeline}, cluster assignments are denoted as $\Phi^{k,c}_{\lambda=l}$, where $k$ is the segment designation (a-e), $c$ is the cluster number (1 or 2), and $l$ is the tip-speed ratio set point. Conditional-averages across $n$, based on cluster assignment, for any variable $X$ (e.g., $\eta$, $\vec{\boldsymbol{\Phi}}$) within one $S_k$ are expressed as $\langle X \vert \Phi^{k,c}_{\lambda=l}\rangle$. In cases where examining individual trajectories of $X$ within a cluster is preferred, an equivalent set notation, $\{ \}$, denotes the subset of cycles of $X$ assigned to the specified cluster. For example, a set of individual $\eta$ trajectories based on $\vec{\boldsymbol{\Phi}}$ clusters is written as 
\begin{equation}
\eta_{\Phi}(\lambda,\theta,n,k,c) = \{\eta(\lambda,\theta,n) \vert \Phi^{k,c}_{\lambda=l} \}.
\label{condperf}
\end{equation}
While the clusters are agnostic to $\eta$, we manually assign cluster 1 (i.e., $c=1$) to the cluster with a higher associated time-averaged performance, $\bar{\eta}$, for full revolutions. To quantify the statistical significance of the identified clusters, we utilized a two-sided Wilcoxon Rank Sum test (\cite{mann1947}, MATLAB ``ranksum'' function) with a 5\% significance level. The null hypothesis is that the populations of $X$ contained in each cluster are drawn from the same continuous distribution and have equal medians. Rejection of the null hypothesis means the populations of $X$ contained in each cluster are statistically significant.

Considering how individual trajectories or cluster conditional-averages differ from the phase-averages over all $n$ can illuminate cluster-specific characteristics. This is done for both performance and flow fields. Conditional performance perturbations, $\eta'_{\Phi}$, are defined as the differences between the individual performance trajectories in each cluster and the phase-averaged performance (average for a single operating condition and azimuthal position across all cycles), $\langle\eta(\lambda,\theta,n)\rangle$, within the same $S_k$. They are computed as 
\begin{equation}
\eta_{\Phi}'(\lambda,\theta,n,k,c) = \eta_{\Phi}(\lambda,\theta,n,c) - \langle\eta(\lambda,\theta,n)\rangle
\label{etaprime}
\end{equation}
These perturbation quantities help highlight where the clusters are performing better or worse than the phase-average.

Conditional difference fields, $\Phi'$, highlight how the conditional-averaged velocity magnitude fields,
$||\langle\vec{\boldsymbol{\Phi}} \vert \Phi^{k,c}_{\lambda=l}\rangle||$, differ from the phase-averaged fields, $||\langle\vec{\boldsymbol{\Phi}}\rangle||$. As with the phase-averaged fields, the conditional-averaged velocity magnitudes are computed from the component averages (i.e., $||\langle u_{rel}\rangle,\langle v_{rel}\rangle||$). The conditional difference fields, computed at each point in space ($x$,$y$), are defined as
\begin{equation}
\Phi'_{\Phi}(\lambda,\theta,k,c,x,y) = ||\langle\vec{\boldsymbol{\Phi}} \vert \Phi^{k,c}_{\lambda=l}\rangle||  - ||\langle\vec{\boldsymbol{\Phi}}\rangle||.
\label{phiprime}
\end{equation}

\section{Results}
\label{results}

We explore the three potential sources of cycle-to-cycle performance variability: variations in near-blade flow fields, variations in inflow velocity, and hysteresis from previous cycles. Section \ref{perfvar} describes the statistical variability in performance and flow fields, highlighting the influence of tip-speed ratio and the blade's phase in the rotation. In Section \ref{flowperfcorr}, we utilize flow-field clustering to correlate cycle-to-cycle performance and flow-field variability. We conclude with a discussion of the impact of the freestream velocity perturbations in Section \ref{freestreamimpact} and hysteresis in Section \ref{nextcycle}.

\subsection{Performance and Flow-field Variability}
\label{perfvar}

The tip-speed ratio affects both time-average performance and the properties of cycle-to-cycle variation around that average, as visualized in Figure \ref{timeaverage}. Mean performance and its variability both increase with tip-speed ratio but performance reaches a maximum near $\lambda\:=\:2.9$, while variability continues to increase at higher tip-speed ratios. For the remainder of this work, we focus on two cases: $\lambda\:=1.5$ (sub-optimal performance) and $\lambda\:=2.5$ (near-optimal performance) which exhibit differences in both time-averaged performance and cycle-to-cycle variability. Both cases are relevant to turbine operation because maximizing efficiency (optimal tip-speed ratio) is a control objective when the freestream velocity is below the turbine's rated speed, while the sub-optimal case corresponds to a strategy for shedding power above the rated speed. We note that the larger variation associated with $\lambda\:=\:2.5$ is due in part to the larger performance and address this later by normalizing by the time-averaged performance as the coefficient of variation. 

\begin{figure*}[h!]
  \centering
    \includegraphics[width=0.9\linewidth]{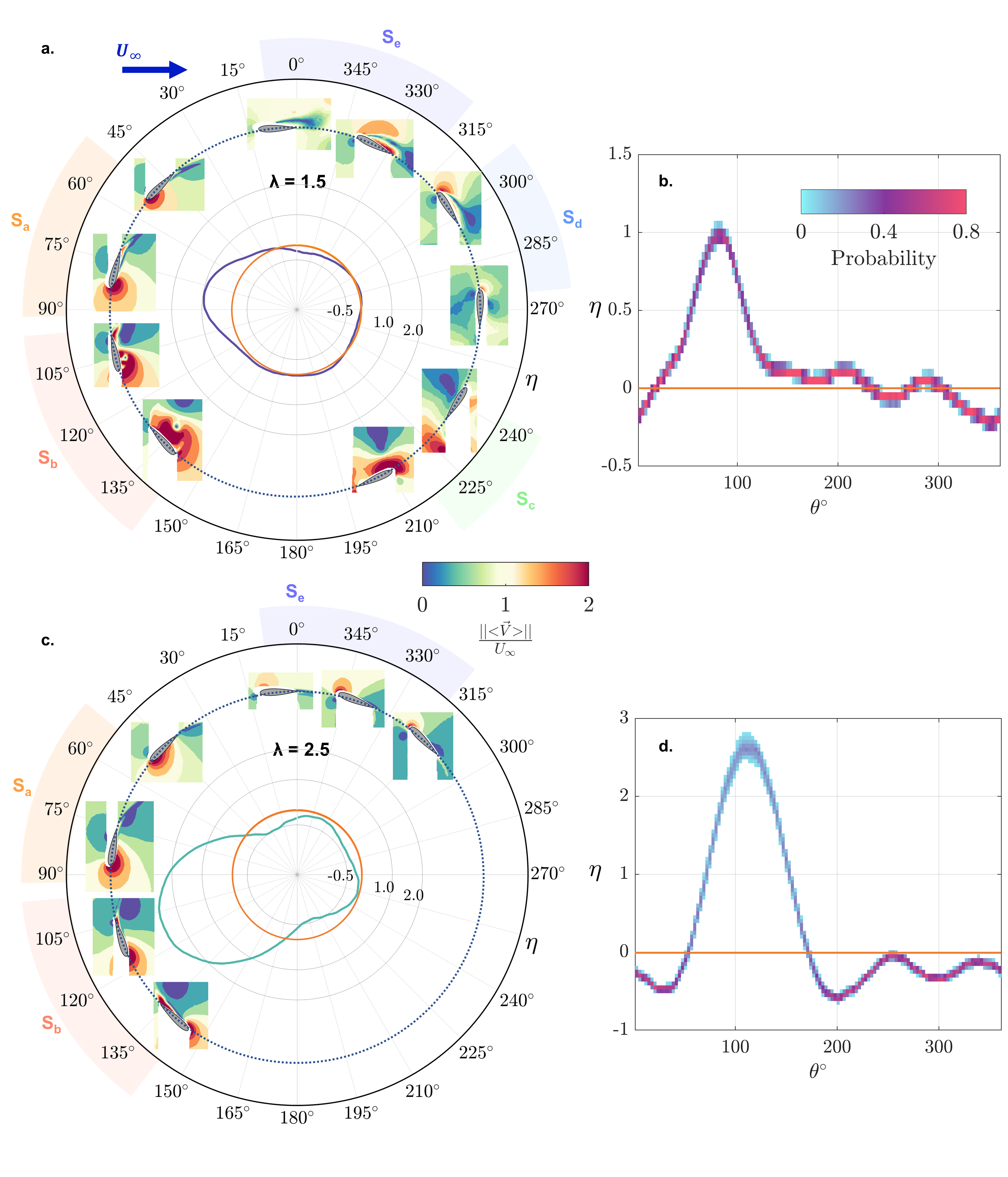}
    \caption{Phase-averaged coefficient of performance for a (a) deep dynamic stall case at $\lambda\:=\:1.5$ and a (c) light dynamic stall case at $\lambda=2.5$. The orange line corresponds to $\eta\:=\:0$, demarcating the power producing from power consumptive phases. Inset are phase-averaged global velocity magnitude fields (normalized by the freestream velocity). Note that the blade and turbine diameter (blue dashed line) are not to scale. The velocity fields are presented in the fixed global reference frame. Shaded regions at the periphery denote the radial extent of flow segments for PIV, $S_k$. (b,d) Bivariate distributions of phase-specific coefficient of performance (blue-to-purple shading with 0.05 $\Delta\eta$ and $2^{\circ}$ $\Delta\theta$).}
    \label{15perfcirc}
\end{figure*}

A defining feature of cross-flow turbine performance is periodic variation within a cycle (i.e., performance variation with $\theta$).  
The polar representations of performance and flow fields for both tip-speed ratios in Figure \ref{15perfcirc} show that the largest cycle-to-cycle variation in phase-specific performance occurs in the upstream sweep around the performance peak  ($\theta\:=\:30^\circ\:-\:135^\circ$ for $\lambda\:=\:1.5$ and $\theta\:=\:60^\circ\:-\:165^\circ$ for $\lambda\:=\:2.5$). In agreement with time-averaged performance (Figure \ref{timeaverage}), the $\lambda\:=\:1.5$ case exhibits less cycle-to-cycle variability during the smaller, narrower performance peak (lower time-averaged performance) than the $\lambda\:=\:2.5$ case. The phase-specific variation in performance is appreciable, but secondary in comparison to the range in performance over a cycle. For both tip-speed ratios, the bivariate distributions of performance are unimodal at each $\theta$, suggesting that this local variation is adequately described by the mean and a descriptor of the spread \citep{Harms}. 

The phase-averaged flow fields of global velocity magnitude around a cross-flow turbine (Figure \ref{15perfcirc}) depend strongly on blade position and tip-speed ratio. Overall, dynamic stall severity and turbine performance are inversely proportional for the two cases (i.e., the higher tip-speed ratio decreases stall severity and increases performance). Specifically, coherent structures from dynamic stall are more apparent for the sub-optimal case, $\lambda\:=\:1.5$, than for the near-optimal case, $\lambda\:=\:2.5$. In comparing the flow fields with turbine performance, we see that for $\lambda\:=\:1.5$, the performance peak coincides with formation, growth, and shedding of the  dynamic stall vortex. On the other hand, for  $\lambda\:=\:2.5$, the flow remains primarily attached until maximum performance, beyond which we observe increasing separation at the trailing edge. For both tip-speed ratios, performance is at a minimum during the downstream sweep, consistent with lower incident velocities due to extraction of momentum from the flow during the ``power stroke''. The flow fields are in qualitative agreement with dynamic stall theory (Section \ref{theory}), as well as prior experiments \citep{sebstalldilema,Simao,Me} and simulation \citep{Coriolis,Mukul} for cross-flow turbines. Because the lower tip-speed ratio leads to a wide oscillation in angle of attack, the $\lambda\:=\:1.5$ case experiences ``deep'' dynamic stall. This is evidenced by a strong dynamic stall vortex that is shed before the maximum nominal angle of attack (Figure \ref{15perfcirc}a), as well as prolonged post-stall flow separation. In contrast, for $\lambda\:=\:2.5$, the foil experiences only ``light'' dynamic stall with smaller vortex structures, limited flow separation until near the maximum nominal angle of attack, and faster post-stall flow recovery. 

\begin{figure*}[t!]
    \centering
    \includegraphics[width=1\linewidth]{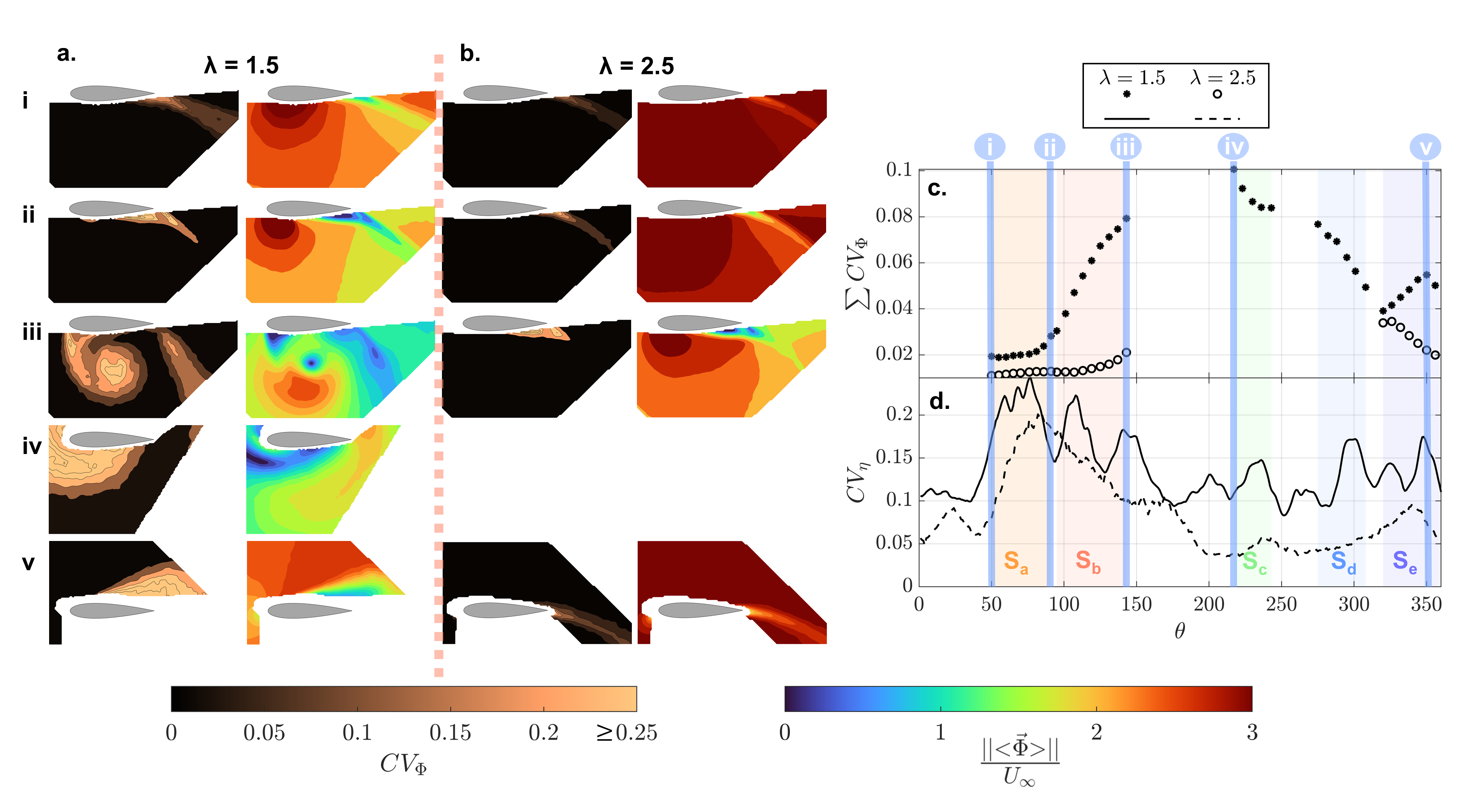}
    \caption{Coefficient of variation fields (colorbar has been truncated at 0.25 for visualization) at select $\theta$ for (a) $\lambda\:=\:1.5$ (b) and $\lambda\:=\:2.5$ and their accompanying relative velocity fields (normalized by the freestream velocity). The location of the $\theta$ positions in (a-b) are labeled in (c-d) which show the (c) sum of the coefficient of variation fields (normalized by the number of data points) (d) and the coefficient of variation for performance for both $\lambda$.}
    \label{perfflowvar}
\end{figure*}

To characterize the cycle-to-cycle variation in the flow fields, we compute the standard deviation fields, $s_{\Phi}$. These are presented normalized by the mean, as the coefficient of variation, to facilitate comparisons with performance variability. Since the flow-field data is sparse in $\theta$, we cannot calculate an accurate time-average, so we define the flow-field coefficient of variation as $CV_\Phi(\lambda,\theta,x,y)=\frac{s_{\Phi}}{\Tilde{\phi}}$, where $\Tilde{\phi}$ is the mean of all the PIV vectors collected at a given phase. Figure \ref{perfflowvar} demonstrates the spatial extent of flow-field variability as a function of $\theta$ and $\lambda$. The flow-field coefficient of variation ranges from $1\%$ to $110\%$ and is truncated in the figure for visualization. During the early portion of the cycle (locations i and ii, $S_a$), variability is confined to a region on the order of the foil thickness close to the blade and in the wake (Figure \ref{perfflowvar}a,b) for both tip-speed ratios. The two cases diverge around $\theta$ = 90$^\circ$ (beginning of $S_b$). There, for $\lambda\:=\:1.5$, flow-field variability expands in the vicinity of the dynamic stall vortex. In contrast, at the same $\theta$, variability for $\lambda\:=\:2.5$ remains confined to the blade wake and the region of separated flow at the trailing edge. By the end of the cycle (location v, $S_e$), for $\lambda\:=\:2.5$, variation is once again primarily limited to the blade wake, while, for $\lambda\:=\:1.5$, we observe high variability in a large region of separated flow (locations iv-v, $S_c - S_e$). We can compactly visualize the trends across phase by summation of ${CV_\Phi}$ in space (i.e., $\sum{CV_\Phi}$). As shown in Figure \ref{perfflowvar}c, $\sum{CV_\Phi}$ is higher at all resolved phases for $\lambda\:=\:1.5$ than $\lambda\:=\:2.5$. For $\lambda\:=\:1.5$, $\sum{CV_\Phi}$ increases dramatically during the growth of the dynamic stall vortex ($S_b$) and, for the resolved phases of the power stroke, reaches a maximum at location iii. This is in agreement with prior work that has consistently demonstrated large increases in cycle-to-cycle variation post-stall \citep{Lennie,Kuppers,Ramasamy2021}. In contrast, for $\lambda\:=\:2.5$, $\sum{CV_\Phi}$ remains relatively low during the power stroke, and the maximum variability likely occurs in a portion of the flow field not visualized in these experiments. For both cases, the highest observed flow-field variability occurs in the downstream sweep. In summary, cycle-to-cycle flow-field variability increases with stall severity and is concentrated around coherent structures originating from the blade or in the blade wake.

The increase in flow-field variability for the lower tip-speed ratio is expected given the higher stall severity, but is inverted from the trends in time-average and phase-specific performance variability, which are higher for $\lambda\:=\:2.5$ (Figure \ref{timeaverage} and \ref{15perfcirc}). However, the coefficient of variation for performance, $CV_\eta=\frac{s_\eta}{\bar{\eta}}$ (Figure \ref{perfflowvar}d), is higher for $\lambda\:=\:1.5$ across most of the cycle, consistent with flow-field variability. In other words, while $\lambda\:=\:2.5$ exhibits higher absolute performance variability, the $\lambda\:=\:1.5$ case has higher relative variability.  

Finally, it is instructive to directly compare the magnitude and timing of flow-field and performance variability. Overall, we observe a non-coincidence between the performance and flow-field variability for both tip-speed ratios. For $\lambda\:=\:1.5$, $\sum{CV_\Phi}$ lags $CV_\eta$, particularly in $S_a$ (Figure \ref{perfflowvar}c,d), which suggests that corresponding flow-field variability present during maximum performance variability likely occurs too close to the blade to be resolved by these PIV measurements. It is also possible that our interpretation is influenced by the $\sum{CV_\Phi}$ formulation which equally weights the flow-field variability regardless of proximity to the blade. For $\lambda\:=\:2.5$, we similarly see the $\sum{CV_\Phi}$ is minimal during the phases of maximum performance variability. For both tip-speed ratios, the observed high flow-field variability in the downstream sweep is associated with lower relative performance variability further in Section \ref{flowperfcorr}).

\begin{figure*}[h!]
    \centering
    \includegraphics[width=1\linewidth]{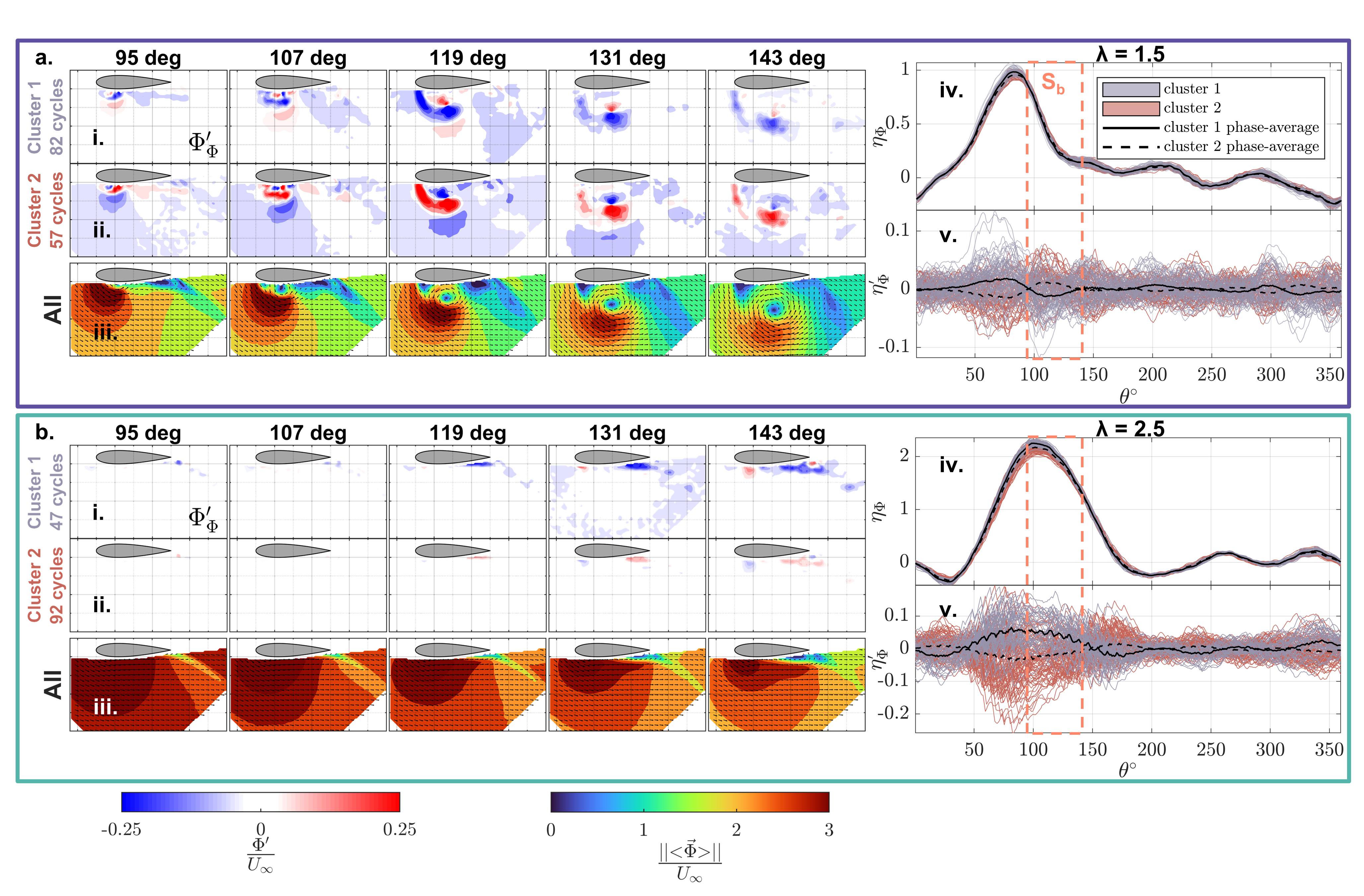}
    \caption{Clustering analysis overview for for (a) $\lambda\:=\:1.5$ (b) and $\lambda\:=\:2.5$. (i,ii) Cluster conditional difference fields and (iii) phase averages for selected $\theta$, with (iv) corresponding performance trajectories and (v) performance perturbations. The grid spacing in (i), (ii) and (iii) is $C/4$. In (iv), each line is colored by cluster assignment and the black lines represent the cluster conditional-averages for performance or performance perturbations (Equations \ref{condperf} and \ref{etaprime}). The dashed rectangles denote the $\theta$ range ($S_b$) where the flow fields in (i),(ii) and (iii) were captured.}
    \label{152_252}
\end{figure*}

\subsection{Flow-field Clusters and Their Associated Performance}
\label{flowperfcorr}

Due to the non-coincidence between the performance and flow-field variability, we must employ a different approach to better understand relationships between the observed performance and flow-field variability. 
By utilizing the flow-field clusters and their associated, cycle-specific performance trajectories, $\eta_{\Phi}$, we are able to characterize the flow-field and performance variability more completely, as well as identify correlations between them. 
For this analysis, we focus on segment $S_b$, which encompasses the end of the power stroke for both tip-speed ratios and the shedding of a dynamic stall vortex for $\lambda\:=\:1.5$. 

Figure \ref{152_252} summarizes the comparison of the flow-field clusters for segment $S_b$ and their corresponding performance trajectories for the two tip-speed ratios. As mentioned in Section \ref{clusterassign}, we designate cluster 1 to be the higher-performing cluster in terms of time-averaged performance. The conditionally-averaged difference fields (i and ii) highlight the deviation between the cluster conditionally-averaged fields and the phase-averaged fields (iii). Animated versions of the phase-averaged flow fields are available in Online Resource 2. The performance trajectories (iv) and their perturbations from the phase-average (v) for each cycle reveal correlations between performance and the flow-field clusters. For both tip-speed ratios, the conditionally-averaged difference fields reveal opposing behaviors between the clusters (e.g., regions of lower-than-average relative velocities in one cluster are coincident with regions of higher-than-average relative velocities in the other). In line with the coefficient of variation fields (Figure \ref{perfflowvar}), the differences are more pronounced for $\lambda\:=\:1.5$ than for $\lambda\:=\:2.5$. Yet, for both tip-speed ratios, performance trajectories are separated by their associated flow-field clusters and exhibit opposing behaviors around the phase average (iv and v). More significantly, the flow-field clusters for both tip-speed ratios are correlated with differences in maximum performance (Figure \ref{152_252}iv), even though $S_b$ does not fully span the performance peak. This suggests that the flow-field variability in $S_b$ stems from variability in the growth and shedding of hydrodynamic structures that originate from the blade, but are unresolved in segment $S_a$ due to their proximity to the surface. Finally, we note that, for $\lambda\:=1.5$, the cluster-averaged performance perturbations reveal that cluster 1 and cluster 2 alternate phases of superior performance over the cycle, even though cluster 1 out-performs cluster 2 on a time-averaged basis.

For brevity, discussion of the relationship between the flow-field clusters in segments other than $S_b$ and their associated performance is presented in Online Resource 1. In brief, despite the apparent contradiction between relatively high flow-field variability and relatively low performance variability in the downstream sweep (Figure \ref{perfflowvar}), the flow-field clusters for segments in the downstream sweep are still connected to time and phase-averaged performance.
 
We now dive deeper into the two tip-speed ratio cases for segment $S_b$ to describe how the flow fields differ between clusters and their relationship with performance. We primarily focus on the $\lambda\:=\:1.5$ case since it exhibits stronger flow-field variability. We then utilize the $\lambda\:=\:2.5$ case to highlight how, despite the lower flow-field variability and limited near-blade resolution, we are still able to extract useful insight from the flow-field clusters. 

\subsubsection{$\lambda\:=\:1.5$}
\label{TSR15}
For this tip-speed ratio, the deep dynamic stall results in growth and shedding of a dynamic stall vortex in segment $S_b$. Difference fields for the cluster with higher time-averaged performance, cluster 1, highlight primarily lower-than-average velocities, with only small regions of higher-than-average velocity. The opposite behavior is seen for the poorer performing cluster, cluster 2. These velocity differences suggest several possible mechanisms: the dynamic stall vorticies for the two clusters differ in strength, in their location with respect to the blade (a consequence of different shedding timing), or both.

\begin{figure}[t!]
    \centering
    \includegraphics[width=0.5\linewidth]{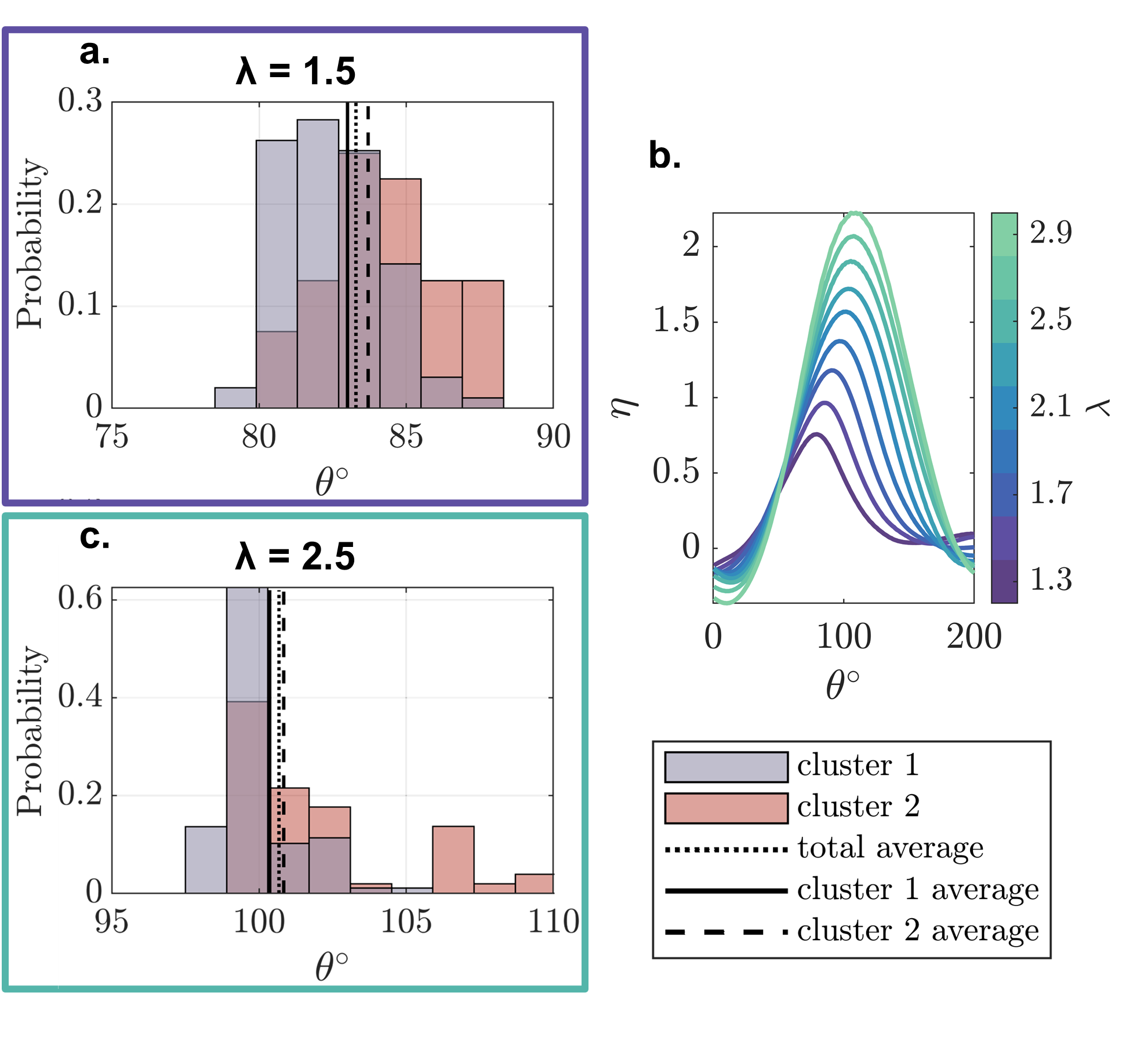}
    \caption{Histograms of cycle-specific phase associated with maximum performance for (a) $\lambda\:=\:1.5$ (c) and $\lambda\:=\:2.5$. (b) Phase-averaged coefficient of performance for select tip-speed ratios.}
    \label{TSRpeaks}
\end{figure}

The difference in time-averaged performance between the two clusters is comparatively small (1.1\%\ with respect to the time-average), but, as apparent in Figure \ref{152_252}a-iv, the performance peak for cluster 1 is slightly higher in amplitude and occurs slightly earlier in the cycle. The shift in phase of maximum performance is highlighted in Figure \ref{TSRpeaks}a. On average, maximum performance occurs at $\theta\:\approx\:83^\circ$ for cycles in cluster 1 and at $\theta\:\approx\:83.9^\circ$ for cycles in cluster 2. The distributions of the phase of maximum performance are statistically significant, as per the rejection of the null hypothesis for the Wilcoxon rank sum test (Section \ref{clusterassign}). These differences in the timing and amplitude of the performance peak between the clusters counter the phase-average trends over all tip-speed ratios (Figure \ref{TSRpeaks}b) where low tip-speed ratios have a performance peak earlier in the cycle and lower maximum and time-averaged performance. 

To gain further insight into how flow fields differ between the two clusters, we first consider modal analysis. A key benefit of the PCA pre-processor is the ability to interpret the resulting clusters through their modes and modal coefficients. 
The directions of the largest variance in the data are described by the modes and sorted by the singular values. Thus the first mode describes the hydrodynamics with the highest variability and the variation described decays as mode number increases. By considering the modal flow fields and their accompanying weighted activation profiles, we can identify the dynamics most associated with the variation and how their contribution changes with blade position and/or between clusters. 

\begin{figure*}[h!]
    \centering
    \includegraphics[width=0.9\linewidth]{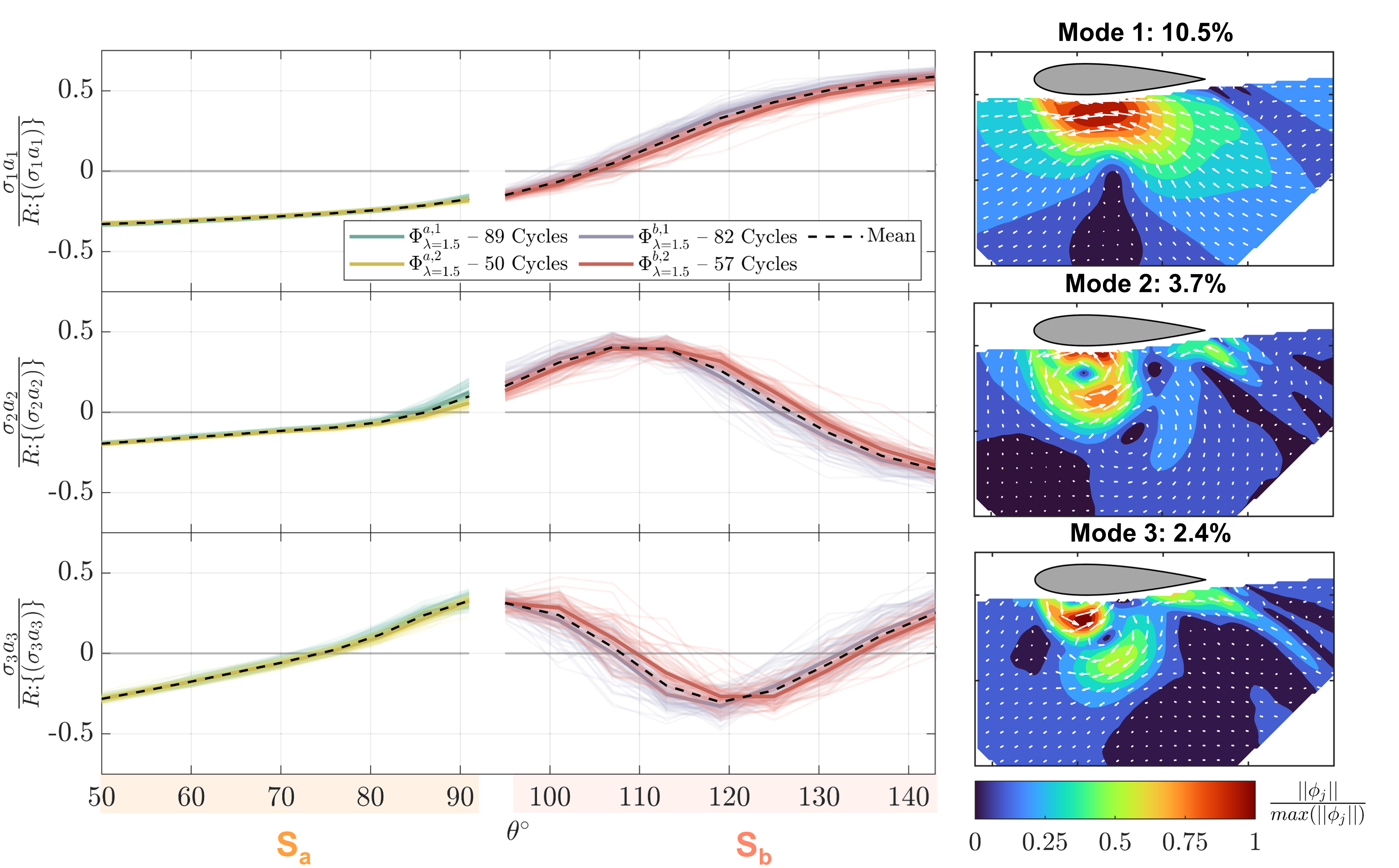}
    \caption{(left) Weighted activation profiles across each cycle, colored by the flow-field cluster assignments for $\lambda\:=\:1.5$ are presented for $S_a$ and $S_b$. The opaque thick lines are the conditional-averages for each cluster and the black dashed line is the phase-average over all cycles. (right) Magnitude fields for the first three modes with variance explained for each mode noted.}
    \label{151 act}
\end{figure*}

Figure \ref{151 act} shows the first three modes and the variability in their weighted activation profiles within segments $S_a$ and $S_b$. These three modes describe $14\%$ of the flow-field variance within these segments (see Online Resource 1 for a plot of the singular values). 
We interpret modes 1 and 2 as related to dynamic stall vortex shedding while mode 3 is attributed to near-blade vortex dynamics occurring earlier in the cycle. In this interpretation, mode 1 represents shedding dynamics that occur once the vortex core is at least a foil thickness away from the blade, while mode 2 occurs earlier in the shedding process. 
The high variance explained by mode 1 is consistent with the velocity magnitude (Figure \ref{perfflowvar}a and \ref{152_252}a-iii) and coefficient of variation fields (Figure \ref{perfflowvar}a), which are dominated by the shedding of the dynamic stall vortex.

We observe that the weighted activation profiles for these three modes are well-converged in $S_a$ but diverge between the flow-field clusters in $S_b$. In $S_b$, the better-performing cluster, cluster 1, has larger weighted activations in mode 1 and the weighted activations are shifted earlier in phase for mode 2. For mode 3, both clusters have similar mean-weighted activations everywhere except for $\theta\:=\:90^\circ\:-\:120^\circ$. This is consistent with low flow-field variability in segment $S_a$ and increased variability in $S_b$ (Figure \ref{perfflowvar}c and Figure \ref{152_252}). At $\theta\:=\:119^\circ$, where the difference fields are most distinct, we see that the higher-than-average velocities closer to the blade for cluster 1 are captured by the higher mode 1 weighted activation. On the other hand, the lower mode 2 weighted activation for cluster 1 at this position captures the lower-than-average velocities farther from the blade. 

\begin{figure}[t!]
    \centering
    \includegraphics[width=0.5\linewidth]{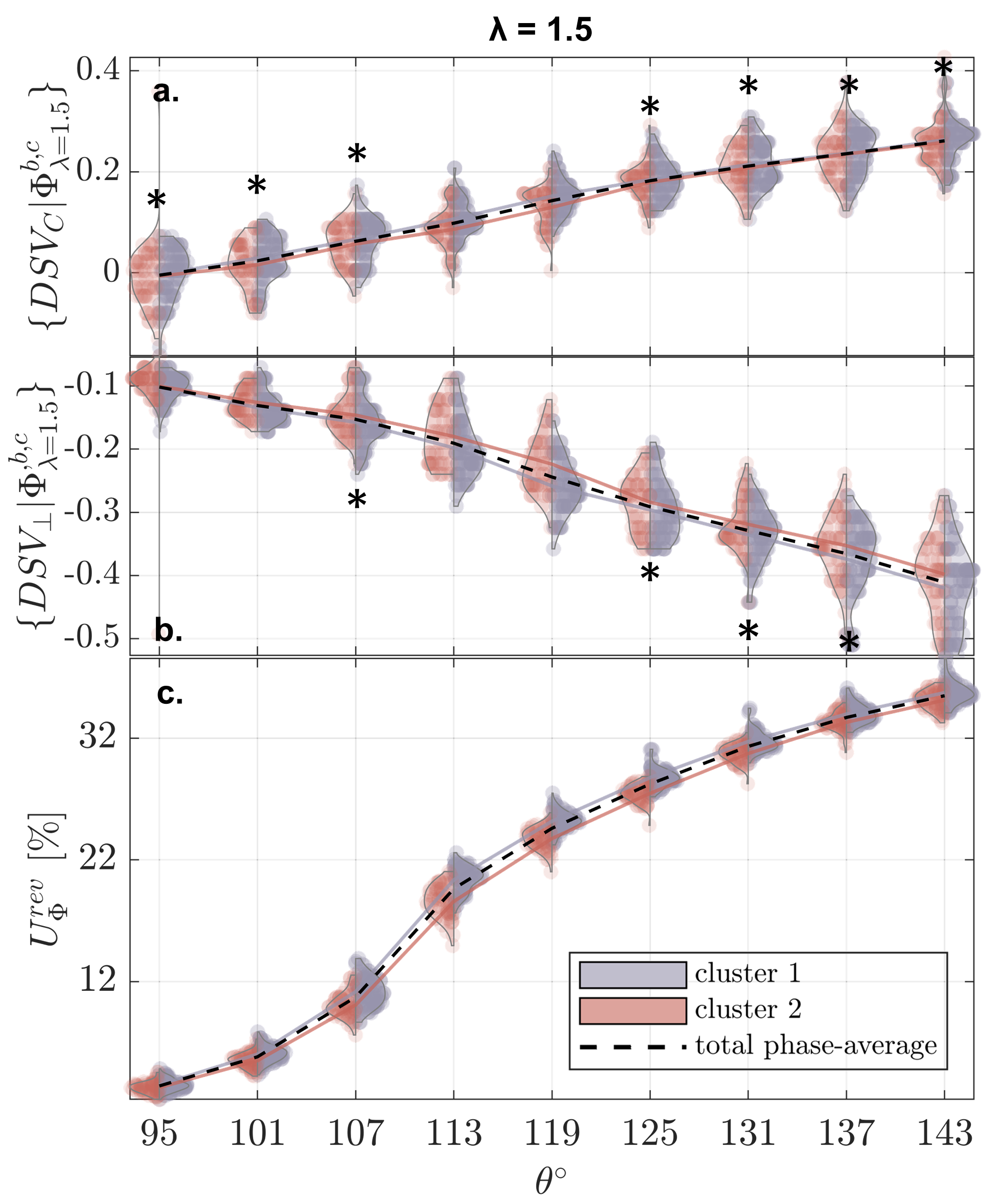}
    \caption{Location of the dynamic stall vortex core in the chord-wise direction, $DSV_C$, (a) and in the chord-normal direction, $DSV_\perp$ (b) for both clusters. Vortex positions are normalized by the chord length, $C$. (c) Reversed flow fraction. The solid lines are the conditional-averages associated with each cluster and the violin plots (\emph{violinplot} MATLAB function from \cite{violin}) at each $\theta$ combine box plots and smoothed histograms to highlight the underlying distribution of the populations. The $\mathbf{*}$ denote phases where the result of the Wilcoxon rank sum test show we cannot reject the null hypothesis that the clusters are samples from continuous distributions with equal medians at the 5\% significance level (i.e., the difference between the two distributions may not be statistically significant).}
    \label{vortrev}
\end{figure}

To establish some physical intuition for the differences between the two clusters in segment $S_b$, we explore the chord-wise and chord-normal position of the dynamic stall vortex, ($DSV_C$ and $DSV_{\perp}$ respectively), and the reversed flow fraction, $U^{rev}_{\Phi}$ (Figure \ref{vortrev}a-c). As we do not have sufficient resolution to resolve the radial extent of the core of a more nuanced vortex model, the position of the vortex core is defined as the location of the maximum value of the swirling strength in individual flow fields reconstructed with 30 PCA modes. The mode truncation removed incoherent noise \citep{PODbasics} resulting in cleaner swirling strength fields that significantly improved tracking performance. 

For $\theta\:=\:113^\circ-125^\circ$, the chord-wise and chord-normal vortex positions are statistically higher for cluster 1 and $DSV_{\perp}$ slope changes earlier for cluster 1. This is indicative of earlier vortex shedding. These results are consistent with \cite{Ramasamy2021} who showed for a pitching foil that clusters based on pressure data were able to reveal earlier shedding of a dynamic stall vortex.

An increase in reversed flow fraction can indicate more flow separation on the blade and therefore more severe stall. We define this quantity as
\begin{equation}
U^{rev}_{\Phi}(\lambda,\theta,n,k,c) = \frac{\sum{}{} [\{u_{rel} \vert \Phi^{k,c}_{\lambda=l}\} \langle 0]}{NxNy}\times100,
\end{equation}
where the $[$ $]$ brackets represent Iverson Bracket Notation which, similar to the Kronecker Delta, returns a value of 1 when the statement in the brackets is true and a value of 0 when false. Cluster 1 exhibits more reversed flow everywhere in $S_b$ (statistically significant). This is consistent with earlier separation and the possibility that cluster 1 cases have stronger dynamic stall vorticies, as evidenced by the higher-than-average velocities near the blade.

In summary, we see clear differences between the clusters for modes that represent various periods of the dynamic stall process, and have shown that the dynamic stall vorticies have different velocity distributions and are shed at different times. Specifically, cycles in cluster 1 have better performance, more reversed flow, higher near-blade velocities in the dynamic stall vortex, and dynamic stall vortex cores that are further from the blade than cluster 2. These hydrodynamics are indicative of an earlier and potentially stronger stall which is consistent with the maximum performance occurring earlier in the cycle (Figure \ref{TSRpeaks}).  

\subsubsection{$\lambda\:=\:2.5$}
\label{TSR25}
The higher tip-speed ratio case presents an opportunity to test the flow-field clustering method when there is less apparent flow-field variability due to the lighter dynamic stall and weaker vortex shedding that remains closer to the blade. Despite this, the difference in time-averaged performance between the two flow-field clusters is 3\% with respect to the total time-average. This is higher than for $\lambda\:=\:1.5$ and consistent with the higher absolute performance variability for $\lambda\:=\:2.5$. Referring once again to Figure \ref{152_252}, the amplitude of the performance peak for the better-performing cluster, cluster 1, is slightly higher, but, unlike for $\lambda\:=\:1.5$, there is no apparent phase shift (Figure \ref{TSRpeaks}c). This is consistent with the convergence in the phase of maximum performance at higher tip-speed ratios in Figure \ref{TSRpeaks}b.  
Despite the relatively low variability, the flow-field clustering results in lower-than-average velocities at the trailing edge of the blade and slightly higher-than-average velocities near the leading edge  for cluster 1, while the opposite is true for cluster 2. 

\begin{figure*}[h!]
    \centering
    \includegraphics[width=0.9\linewidth]{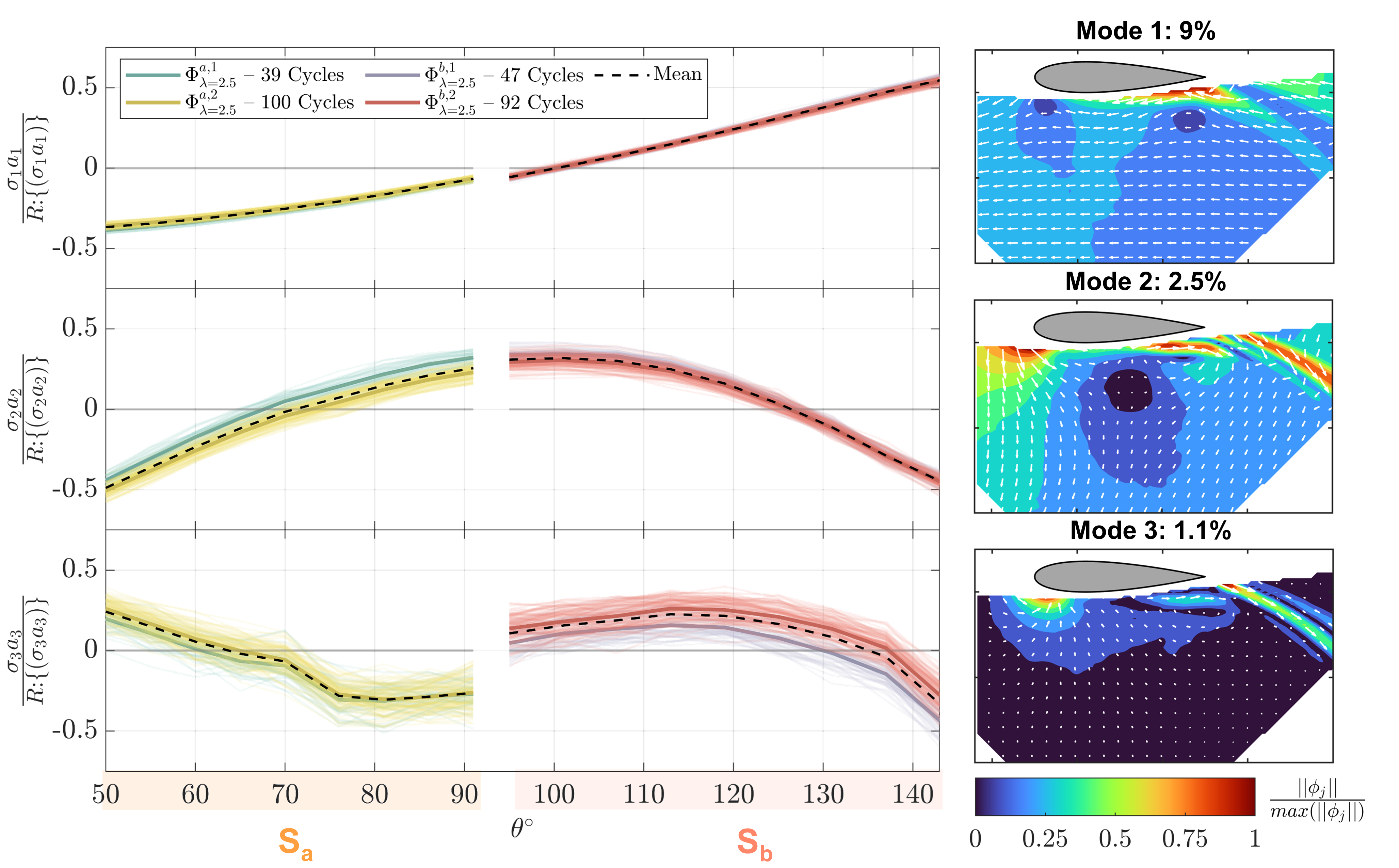}
    \caption{(left) Weighted activation profiles across each cycle, colored by the flow-field cluster assignments for $\lambda\:=\:2.5$ are presented for $S_a$ and $S_b$. The opaque thick lines are the conditional-averages for each cluster and the black dashed line is the phase-average over all cycles. (right) Magnitude fields for the first three modes with variance explained for each mode noted.}
    \label{251 act}
\end{figure*}

The first three modes capture nearly 12\% of the total variance and depict spatial structures on the order of a foil thickness that remain close to the blade and in the wake (Figure \ref{251 act}). We interpret mode 1 as pertaining to attached flow on the blade throughout segment $S_a$, followed by an increase in reversed flow at the trailing edge as $\sigma_1a_1$ crosses zero (representing a change in sign of the mode), 
and the change in direction and magnitude of the relative velocity (Figure \ref{152_252}b-iii). 
Mode 2 represents increased flow along the entire blade when the weighted activations are positive, but 
enhanced separated flow at the trailing edge when negative. 
Similarly, mode 3 describes subtle changes in velocities along the blade. The discontinuity in mode 3 between segments is not likely the result of a rapid change in the dynamics. It is instead most likely a consequence of uncertainty in establishing the blade position between segments, which manifests as a slight misalignment between flow fields of different segments during the translation to the blade-centric reference frame. The $\lambda\:=\:2.5$ case is more sensitive to this error because the energetic dynamics are located adjacent to the blade, and therefore more susceptible to occlusion by the common mask, much more so than for $\lambda\:=\:1.5$. The weighted activation profiles reveal minimal deviations between the clusters, with the exception of mode 2 in $S_a$ and mode 3 in $S_b$. For mode 3, the relatively lower weighted activation for cluster 1 captures the lower-than-average trailing edge velocities. 
These lower velocities are indicative of more flow separation (theoretically limiting lift production) which contradicts the higher time-averaged performance for the cluster. 

In summary, the $\lambda\:=\:2.5$ case exhibits less flow-field variability in comparison to $\lambda\:=\:1.5$, as well as more muted conditional-average difference fields and modes, and less distinct cluster-specific weighted activation profiles. Despite this, the flow-field clusters and their associated performance are distinct and meaningful. 

\subsection{Impact of Freestream Velocity Perturbations}
\label{freestreamimpact} 
\begin{figure}[t!]
    \centering
    \includegraphics[width=0.5\linewidth]{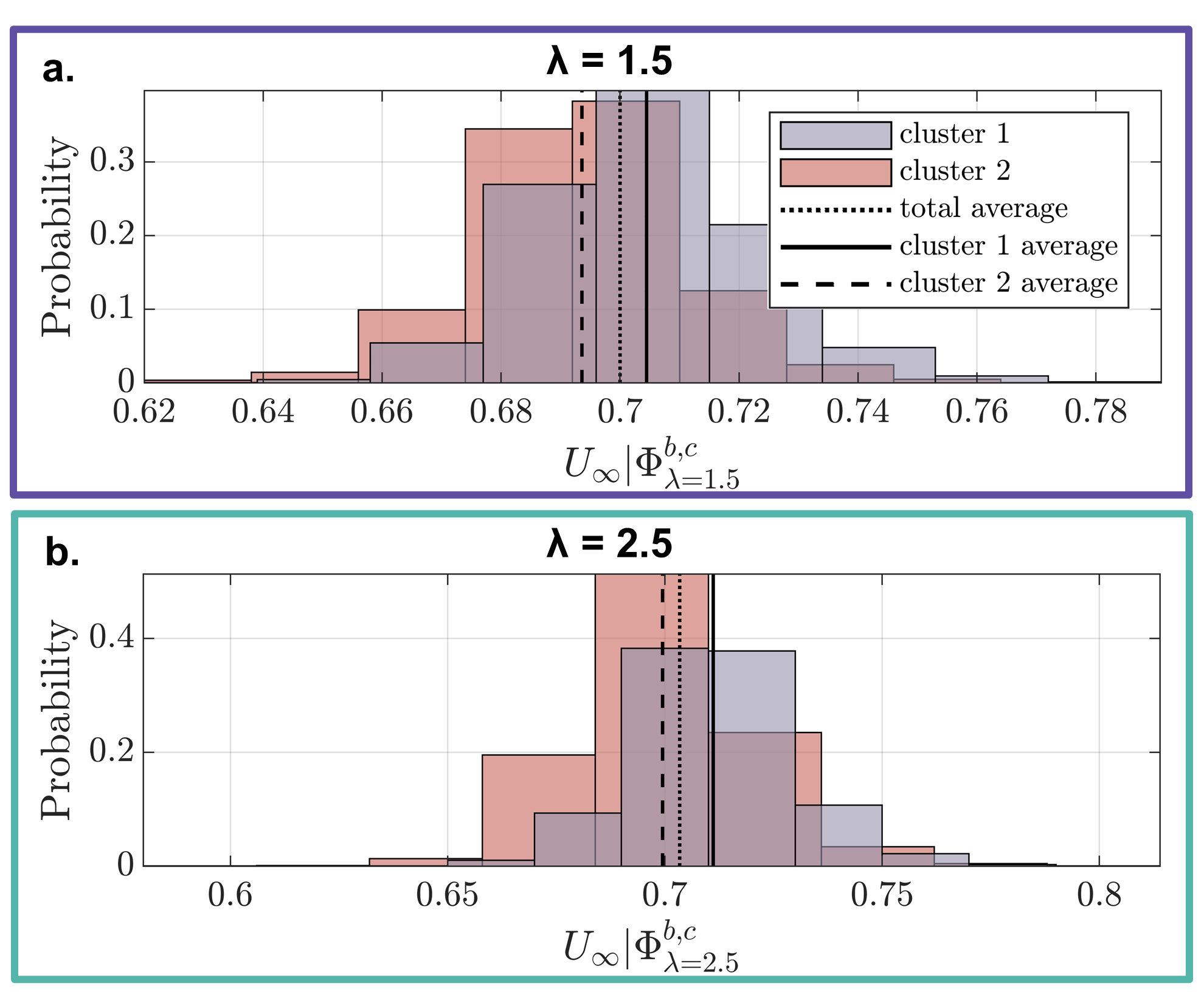}
    \caption{Histogram of freestream velocities for clusters derived from flow fields in $S_b$ at (a) $\lambda\:=\:1.5$ and (b) $\lambda\:=\:2.5$.}
    \label{uinf}
\end{figure}

For both tip-speed ratios, we observe a dichotomy between the flow-field and performance trends. The flow fields for cluster 1 show evidence of earlier and potentially stronger stall, but cluster 1 has higher time-averaged performance. While the flow-field and performance differences between the clusters could be the result of the stochastic dynamic stall process, an alternative hypothesis is that these variations are the result of freestream velocity perturbations. Any changes in the freestream velocity impact both the kinetic energy available in the flow and the instantaneous tip-speed ratio. Cycle-to-cycle variation in the rotation rate is negligible (standard deviations in $\omega$ are 0.02-0.03\% of the time-average), so any perturbations in the freestream will result in cycle-specific tip-speed ratios that differ from the average. Figure \ref{uinf} demonstrates that, for both tip-speed ratios, the instantaneous freestream velocities (advection-corrected as in Section \ref{bladeperfcalc}) are correlated with the flow-field clusters. The difference between the mean of each cluster is 1.6\% and 2.1\% of the time-averaged velocity for $\lambda\:=\:1.5$ and $\lambda\:=\:2.5$, respectively, and are statistically significant. For both tip-speed ratio cases, cluster 1 has a higher conditionally-averaged inflow velocity. As a result, the blade is effectively operating at a lower tip-speed ratio for cluster 1 cycles while also encountering more kinetic energy in the flow. 

\begin{figure}[h!]
    \centering
    \includegraphics[width=0.5\linewidth]{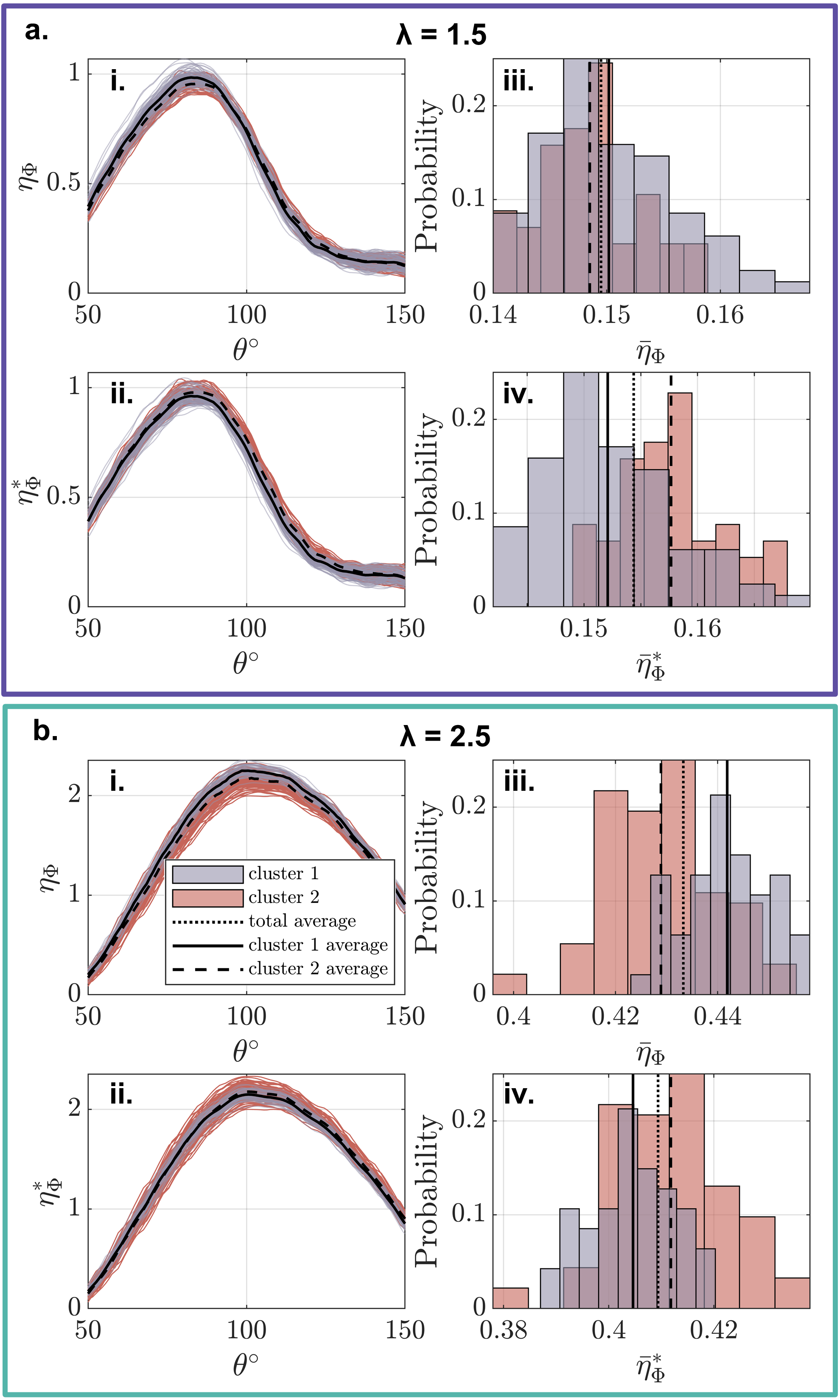}
    \caption{(a) $\lambda\:=\:1.5$ and (b) $\lambda\:=\:2.5$. Individual (i) coefficient of performance and (ii) cluster-specific coefficient of performance trajectories around the peak for the two clusters. The solid lines are the phase-average for cluster 1 and the dashed lines are the phase-average for cluster 2. Cluster-specific histograms of (iii) cycle-specific, time-averaged coefficient of performance and (iv) time-averaged, cluster-specific coefficient of performance.}
    \label{corect}
\end{figure}

Since the coefficient of performance, $\eta_{\Phi}$, is defined using the time-average of the cubed freestream velocity measurements acquired at the tip-speed ratio set point, it does not account for cycle-to-cycle inflow variability. Consequently, we consider a cluster-specific kinetic power, proportional to the conditional-average of the cubed freestream velocities associated with each cluster. Utilizing this, a cluster-specific coefficient of performance, $\eta^*_{\Phi}$, is defined as 
\begin{equation}
\eta^*(\lambda,\theta,n,k,c) = \frac{P}{\frac{1}{2}\rho \overline{\{U_\infty^3\vert\Phi^{k,c}_{\lambda=l}\}}LD}.
\end{equation}
 
A comparison between the performance trajectories around the peak and histograms of cycle-specific, time-averaged performance for $\eta_{\Phi}$ and $\eta^*_{\Phi}$ are presented in Figure \ref{corect}. For $\lambda\:=\:1.5$, both the $\eta_{\Phi}$ and $\eta^*_{\Phi}$ trajectories around the peak are statistically significant between the clusters (Figure \ref{corect}a-i,ii). However, the locally higher $\lambda$ cluster, cluster 2, outperforms cluster 1 at maximum $\eta^*_{\Phi}$. Similarly, the time-averaged distributions for both $\eta_{\Phi}$ and $\eta^*_{\Phi}$ are statistically significant between the clusters, but cluster 2 now outperforms cluster 1. 
These observations are now consistent with phase-averaged (Figure \ref{TSRpeaks}a) and time-averaged  (Figure \ref{timeaverage}) performance trends across tip-speed ratios. 

The results are much the same for the $\lambda\:=\:2.5$ case. In comparison to $\lambda\:=\:1.5$, we observe a better collapse in $\bar{\eta}^*_{\Phi}$ and $\eta^*_{\Phi}$ for the $\lambda\:=\:2.5$ case despite its higher performance variability. This is likely because the angle of attack profiles become less dependent on the tip-speed ratio as the tip-speed ratio increases (Figure \ref{nominal}). This means that, nominally, the sensitivity of the blade kinematics to inflow perturbations is inversely proportional to the tip-speed ratio. As a consequence, the hydrodynamics for $\lambda\:=\:2.5$ are potentially less sensitive to inflow perturbations than for $\lambda\:=\:1.5$.
 
In summary, for both tip-speed ratios, performance based on a cluster-specific freestream kinetic power, $\eta^*_{\Phi}$, confirms the hypothesis that the observed flow-field and performance differences between clusters are primarily caused by inflow velocity variations. This performance dependency on the inflow velocity is in agreement with the axial-flow literature which has consistently shown that turbine power output is correlated with inflow conditions \citep{drualt2020}. Here, the increase in $\eta_{\Phi}$ observed for cluster 1 is a consequence of not accounting for the increased inflow kinetic energy, which increases turbine power output even as the actual efficiency is degraded by the lower cycle-specific tip-speed ratio. While these results suggest that it might be preferable to calculate efficiency on a cycle-specific basis, we found that this approach increases performance variability in this data set. Consequently, we find that cluster-specific performance is more robust to uncertainty in the advection correction (Section \ref{bladeperfcalc}).  

\subsection{Cycle-to-Cycle Hysteresis} 
\label{nextcycle}
\begin{figure*}[h!]
    \centering
    \includegraphics[width=1\linewidth]{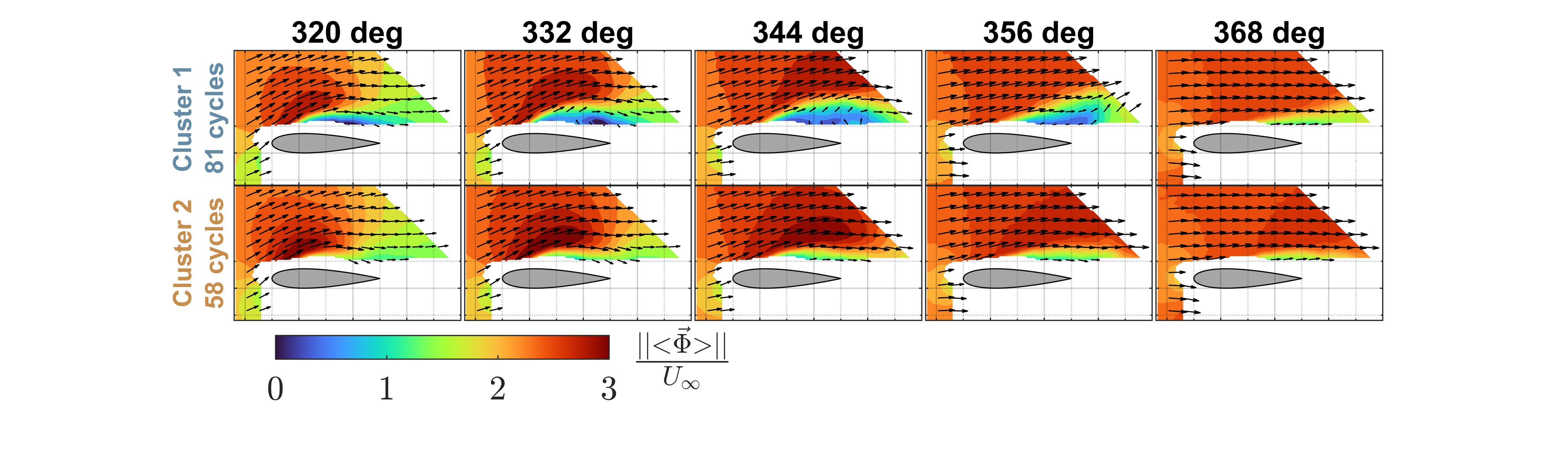}
    \caption{Cluster conditionally-averaged flow fields in $S_e$ for $\lambda\:=\:1.5$ as the blade transitions from the downstream to upstream sweep.}
    \label{153flow}
\end{figure*}
As previously noted (Figure \ref{perfflowvar}c-d, Section \ref{flowperfcorr}), the relatively high cycle-to-cycle flow-field variation during the downstream sweep does not coincide with comparable performance variability. However, it is conceivable that flow-field variation at the end of one cycle could affect future cycles. To explore this possibility, we consider cluster conditionally-averaged flow fields in Figure \ref{153flow}, focusing on $S_e$ for $\lambda\:=\:1.5$,  a case with distinct differences between the two clusters. Flow recovery in cluster 2 appears slower, with more separated flow and lower velocities near the trailing edge at $\theta\:=\:356^\circ$. Despite this, we observe three factors that support a hypothesis of limited hysteresis between cycles. First, as the blade enters the next cycle ($\theta\:=\:368^\circ$), differences between the cluster 1 and cluster 2 flow fields has considerably diminished. 
Second, there is limited cycle-to-cycle performance variability at the beginning of the cycle ($\theta\:<\:40^\circ$,(Figure \ref{perfflowvar}). Third, there is limited variation in the weighted activation profiles for the flow fields at the beginning of the next cycle (Figure \ref{151 act}, $S_a$). If hysteresis from the flow-field variability in the prior cycle was present, we would expect flow-field variability to persist through $\theta\:=0^\circ$ and to see attendant performance variability at the start of the cycle. Thus, it is unlikely that hydrodynamics in the previous cycle contribute to the observed variability in future cycles, especially in comparison to variation associated with the freestream velocity. 

We note that this result is likely influenced by our choice of control scheme. Specifically, if turbine torque, rather than rotation rate, was the regulated quantity, the instantaneous rotation rate would vary within and between cycles, such that the blade kinematics would no longer be deterministic \citep{Briancontrol}. Under such a control scheme, we might see greater variation in the flow fields between cycles and stronger impacts on future cycles.   

\section{Discussion and Conclusion}
\label{discussion}

Cycle-to-cycle performance and flow-field variability for cross-flow turbines are often implicitly neglected through time- and phase-averaging, but, as demonstrated here, are statistically significant. This variability is potentially caused by dynamic stall's stochastic nature, freestream velocity perturbations, and hysteresis from previous cycles. In this work, we explored the extent and sources of cycle-to-cycle variability using near-blade flow fields and performance metrics for sub-optimal, $\lambda\:=\:1.5$, and near-optimal, $\lambda\:=\:2.5$, tip-speed ratios. The flow-field clustering technique developed for this purpose proved sensitive to cycle-to-cycle variations and highlighted correlations between performance and flow-field variability. This technique was effective for both tip-speed ratios despite the lower flow-field variability for $\lambda\:=\:2.5$, where dynamic stall is weaker.   

Overall, performance and hydrodynamic variability were found to be non-coincident in phase and differ in magnitude. Across all phases, the coefficient of variation in performance ranges from $4\:-\:20\%$. While an imperfect comparison, the phase-specific flow-field coefficients of variation are as high as $110\%$, an order of magnitude greater. Performance variability is highest around the maximum performance within a cycle, but, because of limits to near-blade flow-field resolution, observable flow-field variability is minimal until beyond maximum performance. For $\lambda\:=\:1.5$, flow-field variability increases during the growth and shedding of the large dynamic stall vortex, and, for $\lambda\:=\:2.5$, during the growth of the separated flow region at the trailing edge. Despite this lag in the flow-field variability, clusters based on observed flow fields throughout the rotation are correlated with time- and phase-averaged performance for both tip-speed ratios. This is not evident when considering only aggregate, statistical measures of variability. 

Hysteresis and dynamic stall stochasticity may contribute to variability, but freestream velocity perturbations dominate the observed cycle-to-cycle variation in both the flow fields and performance. Given that the high flow-field variability at the end of the turbine rotation is not accompanied by high variability at the beginning of the rotation, it is unlikely that hysteresis impacts are substantial. While we cannot completely disentangle the impacts of freestream velocity perturbations and dynamic stall stochasticity, the cycle-specific freestream velocities are correlated with the flow-field clusters. These velocity perturbations are shown to impact the kinetic energy available in the flow and to perturb $\lambda$ enough to influence the dynamic stall process. For $\lambda\:=\:1.5$, the better-performing cluster for $\theta = 95\:-\:143^\circ$ exhibits higher inflow velocities, a higher reversed flow fraction, earlier dynamic stall vortex shedding, and a performance peak occurring slightly earlier in the cycle. All of these behaviors are consistent with the locally lower $\lambda$, however, the higher performance is counter to the trend where a lower $\lambda$ is associated with lower maximum and time-averaged performance. This contradiction is also present for the $\lambda\:=\:2.5$ case. When performance is calculated with a cluster-specific kinetic power, the apparent contradiction is resolved and the locally higher tip-speed ratio cluster outperforms the other. This suggests, for these conditions, the change in the available kinetic power in the inflow has a greater influence than the perturbation to the local tip-speed ratio. 

A clear performance and hydrodynamic dependence on assigned cluster is observed, despite the freestream turbulence intensity of 1.8-2.1\%, which is relatively low in comparison to field conditions \cite{thomson2012}. That said, the differences in time-averaged performance between clusters are small (1-3\% relative to the time-average of all data). Therefore, for these conditions, phase-averaging flow fields and performance is an effective way to investigate general trends. This work demonstrates that clustering is useful for more nuanced analyses that seek to understand the connections between observed flow fields and turbine performance. In field settings or laboratory conditions with higher turbulence, or the use of torque control, cycle-specific tip-speed ratios could be perturbed further from the phase-average. For such conditions, performance and hydrodynamic variability may increase, making a cluster-specific analysis a more necessary alternative to phase-averaging. Further, the specific mechanisms underlying cycle-to-cycle variability may change for different inflow conditions (e.g., Reynolds number), turbine geometry, and kinematics, as well as for other non-turbine systems experiencing dynamic stall.  

We must note that, in future studies, it would be important to consider the number of clusters as a free parameter. Here, two clusters proved appropriate, but this may not be optimal for all cases. For the current data set, two clusters produce unique phase-averaged weighted activation profiles in the principle component analysis. While we utilize clusters based on the flow fields, PIV data collection is time intensive and generally confined to laboratory settings. Investigating the benefits of clustering on the basis of performance or distributed loads could be a direction for future research. 

The flow-field clustering approach contributes to our understanding of the mechanisms responsible for the performance and hydrodynamic variability of cross-flow turbines. It provides a more comprehensive picture of the phase-varying flow fields than aggregate, statistical representations, and provides conditionally averaged groups that are not based on subjective, hand-engineered metrics. We observe physically meaningful clusters representing a series of distinct flow-field evolutions. These clusters are correlated with performance, and reveal variations in timing of the dynamic stall process.

\section{Acknowledgments}

 The authors thank the Alice C. Tyler Charitable Trust for supporting the research facility and acknowledge the substantial contributions by Benjamin Strom, Hannah Ross, Aidan Hunt, Carl Stringer, Erik Skeel, and Craig Hill for their contributions to the development and upgrades of the experimental setup and code base. We also thank our colleagues in the Pacific Marine Energy Center for their continued support.

\section*{Declarations}

\subsection*{Funding}
Financial support was received from the United States Department of Defense Naval Facilities Engineering Systems Command and through the National Science Foundation Graduate Research Fellowship Program. 
\subsection*{Competing interests}
The authors have no relevant financial or non-financial interests to disclose.
\subsection*{Ethics approval}
Not applicable
\subsection*{Consent to participate}
Not applicable
\subsection*{Consent for publication}
Not applicable
\subsection*{Availability of data and materials}
Data and processing codes are available upon request.
\subsection*{Code availability}
Not applicable
\subsection*{Authors contributions}
All authors contributed to the study conception and design. Material preparation and data collection were performed by Abigale Snortland and Isabel Scherl. All data analysis and code base development was performed by Abigale Snortland. The first draft of the manuscript was written by Abigale Snortland. Brian Polagye and Owen Williams advised throughout this project and were closely involved in the framing and writing of the final manuscript. All authors read and approved the final manuscript.

\begin{appendices}
\section{Performance and Flow-field Uncertainty analysis}\label{sec:Uncertainty}
The presented results involve relatively subtle, but statistically significant, observations in the performance and flow field data. Despite the uncertainties in both data sets, we find statistically meaningful correlations between the different data streams that are in line with theory and current understanding (e.g., statistically meaningful correlations between the flow-field clusters obtained through PIV and measured performance).

\subsection{Performance Uncertainty}

\begin{table*}
\begin{minipage}{1\textwidth}
\label{uncertainty specs}
\caption{Instrument uncertainty for performance testing}
\begin{tabular}{ |c||c|c|l| }
\hline
\textbf{Instrument} &\textbf{ $C_P$ Component} & \textbf{Unit} & \textbf{Systematic Uncertainty} \\
\hline
Nortek Vectrino ADV & $U_\infty^3$ & m/s & Sensor Accuracy: $\pm 1\%$ of measured values $\pm 1$ mm/s \footnote{$U_\infty$ systematic uncertainty converted to $U_\infty^3$ systematic uncertainty by converting the listed values to a percentage of $U_\infty$ and multiplying by 3} \\
\hline
Motor Encoder & $\theta$ & degrees & Sensor accuracy: $\pm 0.004$ \\
\hline
Futek FSH02595 & $\tau$ & N-m & Sensor non-linearity: $0.2\%$ of rated load \\ 
& & & Sensor hysteresis: $\pm 0.2\%$ of rated load \\
& & & Sensor non-repeatability: $\pm 0.05\%$ of rated load \footnote{Summed as root sum of square, rated output of 11.3 N-m yields $\pm 0.0325$ N-m} \\
\hline
\end{tabular}
\end{minipage}
\end{table*}

\begin{figure*}[h!]
    \centering
    \includegraphics[width=1\linewidth]{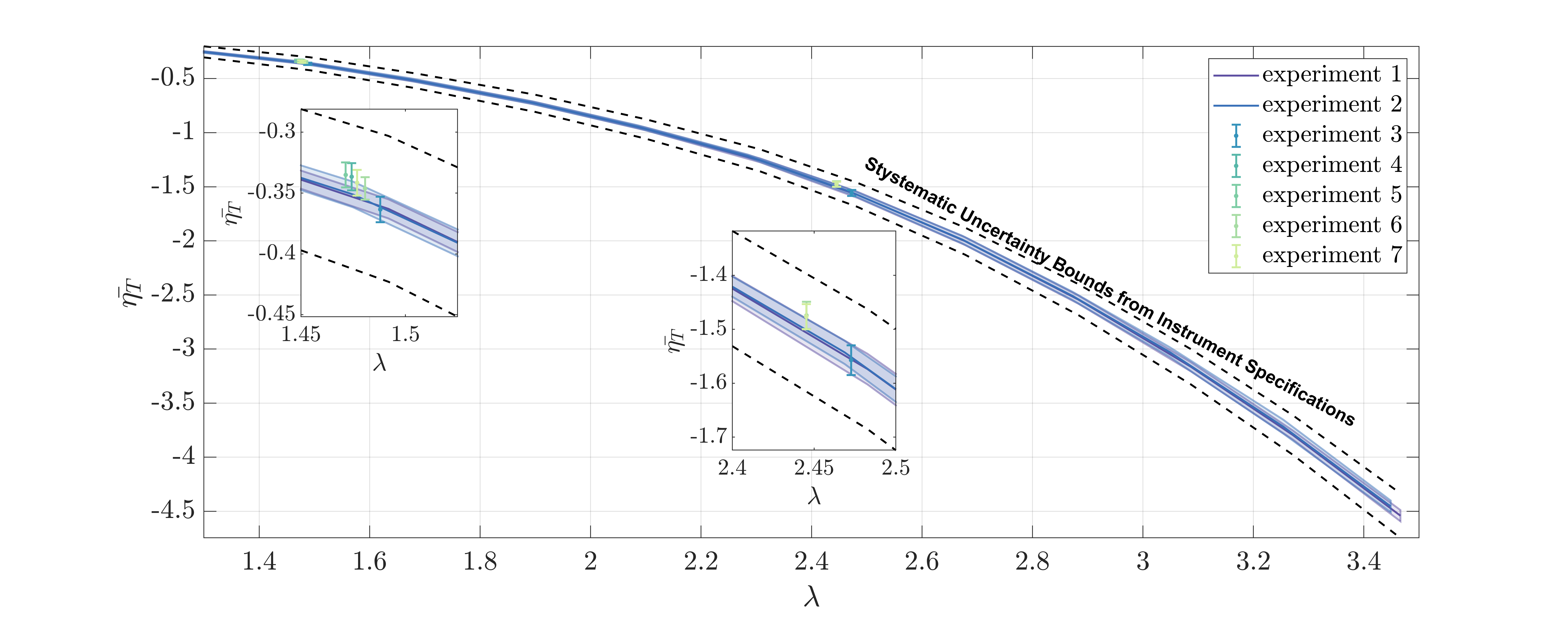}
    \caption{Time-averaged full turbine performance measurements. The shaded regions and error bars represent $\pm$ two standard deviations in the cycle-averaged performance distributions. The dashed black lines represent the systematic uncertainty bounds computed following the methodology in \cite{Cavanaro2016}. The error bars for experiments 3-7 are best shown in the insets.}
    \label{uncertaintycascade}
\end{figure*}

One approach to assessing uncertainty is a formal analysis based on instrument accuracy. However, in performing such an analysis, two issues are encountered. First, the cycle-to-cycle variation that is the focus of this work would be treated as a random uncertainty in a formal analysis. Second, as subsequently shown, the systematic uncertainty indicated by a formal analysis exceeds the experimentally-observed variation.

Cross-flow turbine performance is a function of the measured hydrodynamic torque (Futek FSH02595 torque cell), the rotation rate (derivative of the measured position from the motor encoder), and the freestream velocity (Nortek Vectrino ADV). These measurements are all subject to systematic and random uncertainties. Systematic uncertainty describes the ability of the instrumentation to accurately determine the central moment, while random uncertainties are related to the variability around the central moment. 

The systematic uncertainty for $\eta_T$ (full turbine performance) is quantified following the methodology in Appendix A of \cite{Cavanaro2016}. The systematic uncertainties, as taken from the manufacturer specifications, are listed in Table \ref{uncertainty specs}, and these bounds are shown relative to experimental performance for the full turbine in Figure \ref{uncertaintycascade}. Experimental repeatability is excellent, as evidenced by the small variations in cycle-averaged performance within and across repeated performance measurements. Further, the systematic uncertainty estimated from the instrumentation specifications exceeds the observed variability. We note that the uncertainty analysis is done for the full turbine efficiency. It would be inappropriate to perform this analysis on the blade-level efficiency since the strut subtraction methodology involves independent measurements of the full turbine and the struts in isolation.

 \begin{figure*}[h!]
    \centering
    \includegraphics[width=1\linewidth]{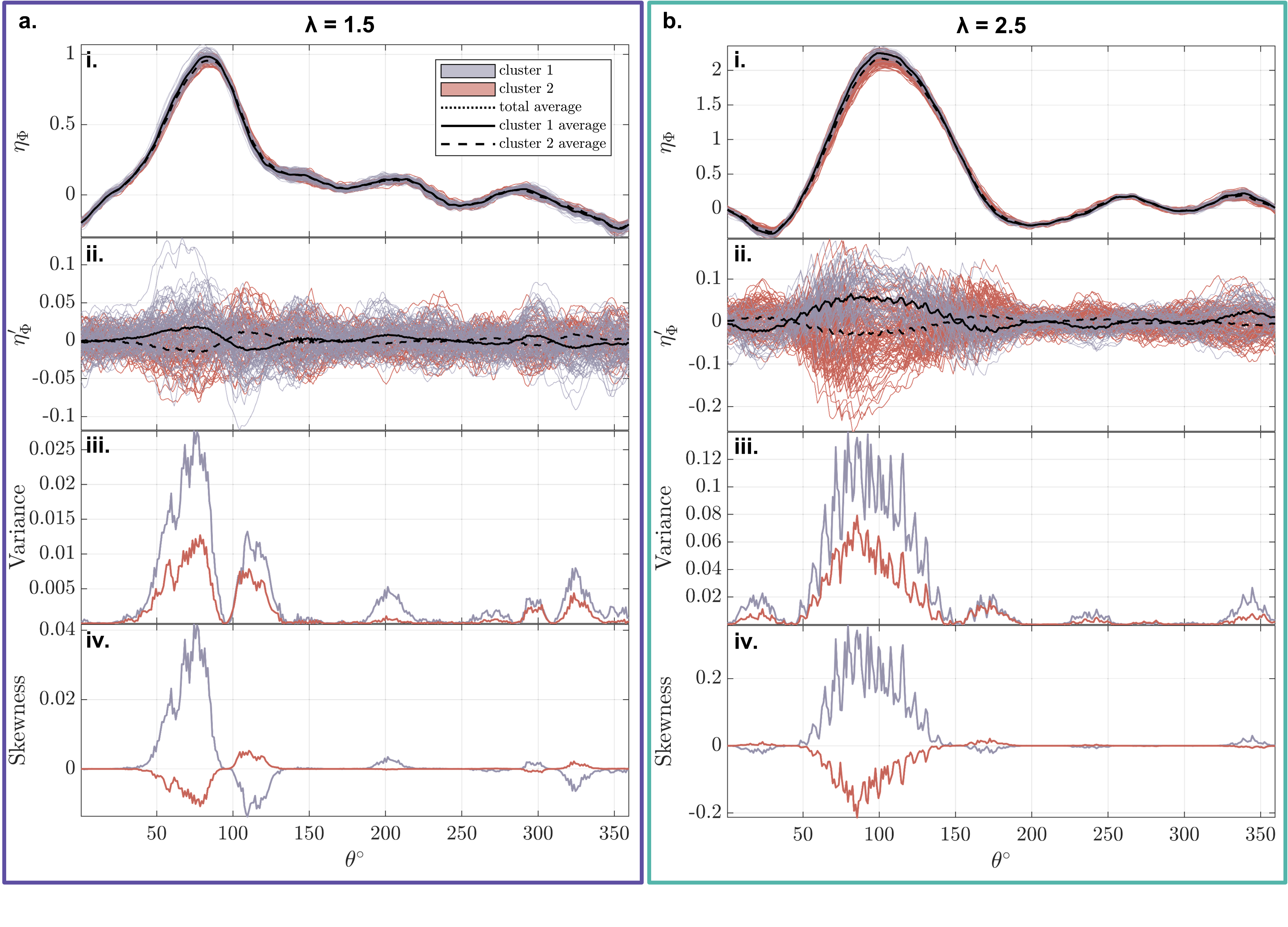}
    \caption{Statitical analysis of the performance trajectories associated with the flow-field clusters for (a) $\lambda\:=\:1.5$ and (b) $\lambda\:=\:2.5$.(i) Performance trajectories and (ii) performance perturbations. The black lines represent the cluster conditional-averages for performance or performance perturbations (Equations \ref{condperf} and \ref{etaprime}). (iii) Variance (Equation \ref{vareqn}) in the performance trajectories associated with each cluster. (Bottom) Skewness (Equation \ref{skew}) for each cluster relative to the phase-average of all cycles.}.
    \label{expandedperf}
\end{figure*}

Given that we are less concerned with the absolute accuracy of the central moment of the distribution given by the systematic uncertainty, it is more instructive to consider the characteristics of the variability in blade-level efficiency around the central moment. The phase-averaged performance perturbations are defined as the difference between performance in a specific cycle relative to the phase-average of all cycles, so the variance (second moment) of the performance perturbations in each cluster is given as
\begin{equation}
\alpha_2 = \frac{1}{n}\sum(\eta')^2.
\label{vareqn}
\end{equation}
Substantial differences in the variance of the two clusters would suggest that the two distributions have different fundamental structure and that a comparison of their mean values (first moment) could be misleading, as well as invalidate the use of the Wilcoxon rank sum test used in Sections \ref{TSR15}-\ref{freestreamimpact}.  

Similarly, the skewness (third moment) is given as
\begin{equation}
\alpha_3=\frac{1}{n}\sum(\eta')^3.  
\label{skew}
\end{equation} 
A value of zero skewness would signify that the performance trajectories in a cluster are normally distributed around the phase-average of all cycles (i.e., cluster trajectories are randomly distributed). Conversely, positive values would signify that performance within a cluster is skewed above the phase-average and vice-versa for negative values.

Figure \ref{expandedperf} presents these moments for both tip-speed ratios. Figure \ref{expandedperf}i-ii shows blade-level performance and performance perturbations for all trajectories (as presented in Figure \ref{152_252}). Here, the black lines denote the first moment of the distribution, emphasizing the relatively subtle trends in the underlying distribution of trajectories assigned to each cluster. We observe that the phase-average variance trends for each cluster are similar (Figure \ref{expandedperf}iii), which shows that the cycles in each cluster are similarly dispersed with respect to the phase-average of all cycles. This validates the use of the Wilcoxon rank sum test and comparisons between the phase-averages are meaningful. However, because variance is always positive, this does not show how the clusters are distributed with respect to the phase-average of all cycles. For both tip-speed ratios, the skewness (Figure \ref{expandedperf}iv) shows that the performance trajectories associated with each cluster are not randomly distributed around the phase-average of all cycles and that they are skewed in opposing directions. If the cycle-to-cycle variability was all attributed to random uncertainty, the observed correlations between the performance and the flow-field clusters would not be present. While random uncertainty is certainly still present, the results presented in this work provide confidence that the cycle-to-cycle variation is driven by physical processes (i.e., inflow perturbations).



  \begin{figure*}[h!]
    \centering
    \includegraphics[width=0.9\linewidth]{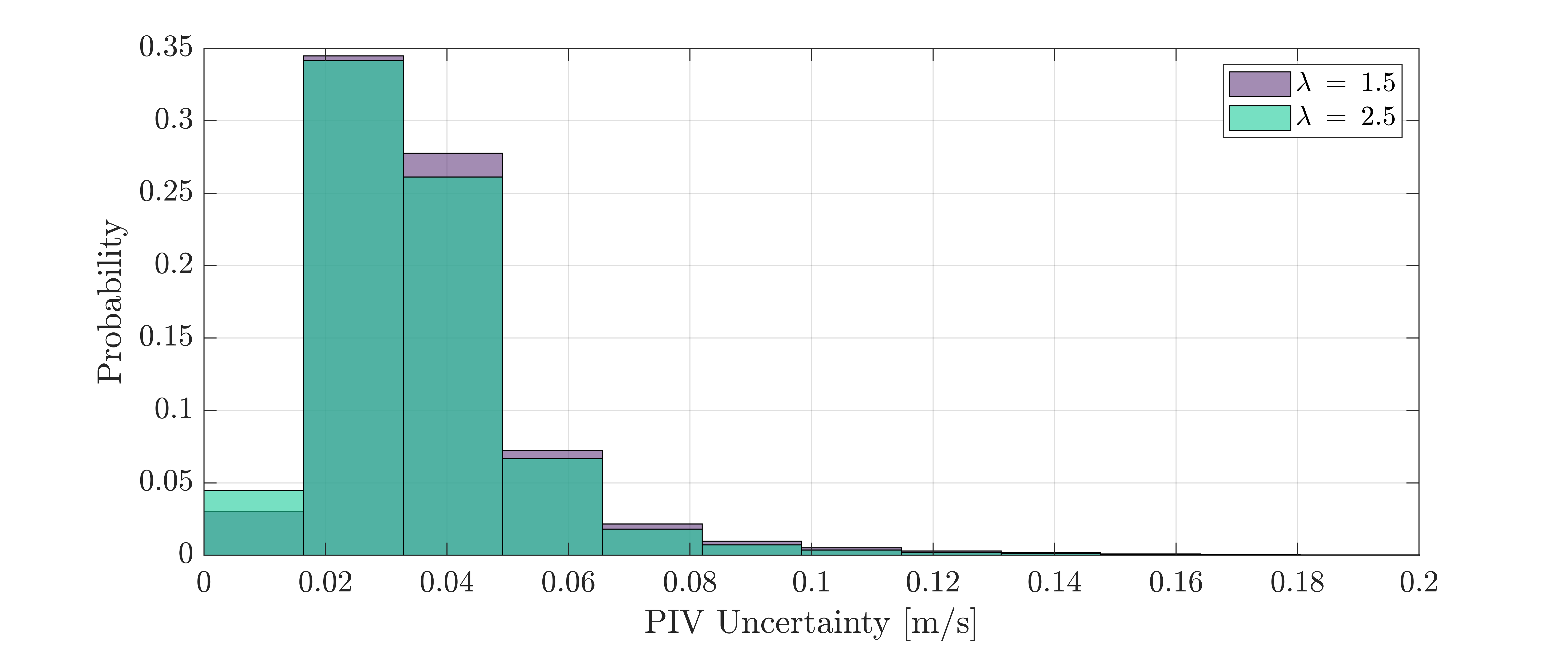}
    \caption{Histograms PIV uncertainty for every vector across all phases for $\lambda\:=\:1.5$ and $\lambda\:=\:2.5$.}.
    \label{uncertainty hist}
\end{figure*}

 \subsection{PIV Uncertainty}
 The PIV uncertainty is quantified by the correlation method employed in LaVision DaVis \citep{PIVuncertainty} which analyses individual pixel contributions to the correlation peak. The output is an uncertainty field corresponding to each individual flow field  (i.e N=139 uncertainty fields at each phase corresponding to the N=139 flow fields at each phase). Histograms of PIV uncertainty for every vector across all phases are given in Figure \ref{uncertainty hist}. The PIV uncertainty is $\pm0.037$ m/s on average (average of all vectors across all phases for a specific tip-speed ratio) for $\lambda\:=\:1.5$ and $\pm0.035$ m/s for $\lambda\:=\:2.5$. These uncertainties are an order of magnitude smaller than the cycle-to-cycle variability observed in the flow fields (Figure \ref{152_252}).

\end{appendices}


\bibliography{References.bib}



\end{document}